\documentclass[fleqn,10pt]{wlscirep}
\usepackage[utf8]{inputenc}
\usepackage[T1]{fontenc}
\usepackage{subcaption}
\usepackage{siunitx}
\usepackage{multirow,hyperref}
\usepackage{soul}
\usepackage{float}
\usepackage{pdfpages}
\usepackage{lastpage}
\usepackage{fancyhdr}
\usepackage{lipsum}

\title{Enhancing the efficiency of quantum-dot-based single-photon source designs by suppressing background emission using concentric rings}

\author[1,*]{Martin Arentoft Jacobsen}
\author[1]{Luca Vannucci}
\author[2]{Julien Claudon}
\author[2]{Jean-Michel Gérard}
\author[1]{Niels Gregersen}
\affil[1]{DTU Electro, Department of Electrical and Photonics Engineering, Technical University of Denmark, DK-2800 Kongens Lyngby, Denmark.}
\affil[2]{Univ. Grenoble Alpes, CEA, Grenoble INP, IRIG, PHELIQS, “Nanophysique et Semiconducteurs” Group, F-38000 Grenoble, France.}
\affil[*]{maaja@dtu.dk}

\begin{abstract}
In this paper, we theoretically demonstrate that a few-period circular Bragg reflector consisting of concentric rings placed around an infinite nanowire with an embedded quantum dot can increase the fraction of radiative emission into the fundamental $\mathrm{HE}_{11}$ mode ($\beta=\Gamma_{\rm HE_{11}}/\Gamma_{\rm Total}$) up to 0.999 due to enhanced suppression of the emission into radiation modes caused by a photonic bandgap effect. We then apply this strategy in the practically relevant case of the finite-sized single-photon source based on tapered nanowires and demonstrate that the collection efficiency can be improved. Additionally, we also show the beneficial effects of placing optimized rings around the micropillar single-photon source.
\end{abstract}
\begin{document}

\flushbottom
\maketitle
%
%
\thispagestyle{empty}


\section*{Introduction}

The quantum optics community is in hot pursuit of designing, developing, and fabricating deterministic sources of single indistinguishable photons\cite{Arakawa2020, Aharonovich2016, Gregersen2013, Gregersen2017} and entangled photon pairs\cite{Huber2018}, since such sources are crucial key building blocks in scalable quantum technologies\cite{Pan2012,OBrien2009}. One of the main figures of merit of the single-photon source (SPS) is the collection efficiency $\varepsilon$, defined as the number of collected photons in the desired out-coupling channel relative to the number of emitted photons from the source. The success probability $P$ of an $N$-photon interference experiment  scales as $P\propto\varepsilon^N$, and it is thus of great importance to increase $\varepsilon$ towards unity.

A main SPS engineering platform is based on a quantum dot (QD)\cite{Gerard1999,Michler2000,Santori2002,Arakawa2020,Aharonovich2016,Shields2007} embedded in a carefully engineered photonic nanostructure\cite{Barnes2002,Gregersen2013,Gregersen2017}. The discrete energy levels of the QD and the large magnitude of the Coulomb interaction between trapped charge carriers allow for deterministic emission of single photons through the spontaneous emission (SE) process. At the same time, the photonic nanostructure is designed to optimize the collection of the photons: The cavity quantum electrodynamics (CQED) design strategy consists of placing the QD in a cavity and exploiting the Purcell effect\cite{JMG1998} in the weak coupling regime to enhance the emission into the optical mode of interest, typically the fundamental cavity mode. The enhancement is quantified by the Purcell factor\cite{Purcell1946} $F_{\rm p}=\Gamma_{\rm C}/\Gamma_0$, where $\Gamma_{\rm C}$ is the SE rate into the cavity mode, and $\Gamma_0$ is the SE rate in a bulk medium. The $\beta$ factor then quantifies how much of the total emission is funneled into the cavity mode\cite{Barnes2002} as
\begin{equation}
    \beta=\frac{\Gamma_{\rm C}}{\Gamma_{\rm T}}=\frac{F_{\rm p}}{F_{\rm p}+\Gamma_{\rm B}/\Gamma_{0}},
    \label{betacav}
\end{equation} where $\Gamma_{\rm T}$ is the total emission rate, and $\Gamma_{\rm B}$ is the SE rate into all other background modes.

 For SPS designs with large Purcell enhancement, the true efficiency $\varepsilon$ is often well described by the single-mode model\cite{Friedler2009, Gregersen:10, Claudon2010, Claudon13, Munsch2013,  Wang2020_PRB_Biying} (SMM) given by $\varepsilon\approx\varepsilon_{\rm SMM}=\gamma\beta$, where $\gamma$ (the transmission coefficient) is the ratio between the collected light of the cavity mode in the out-coupling channel and the emission into the cavity mode. It is thus equally important that $\gamma$ approaches unity to ensure high efficiency. Relying solely on Purcell enhancement to obtain good collection efficiency results in narrowband designs, requiring careful spectral alignment between the QD and the cavity\cite{Somaschi2016}. Still, this approach has so far resulted in the most successful SPS designs, such as the micropillar cavity\cite{Moreau2001,Wang2019b, Wang2020_PRB_Biying, Somaschi2016, Ding2016} and the open cavity geometry\cite{Tomm2021}, demonstrating up to $\varepsilon \sim$~0.6 into a first lens \cite{Wang2019b} and into a fiber \cite{Tomm2021}, respectively, combined with highly indistinguishable photon emission.

Using CQED to increase the Purcell factor is typically also beneficial for the indistinguishability, as the emission rate into the zero-phonon line is strongly increased relative to that through the phonon sideband, which is present due to the solid-state environment. In addition, a global acceleration of SE strongly reduces the impact of pure dephasing. Theoretically, this approach can increase the indistinguishability
towards unity. However, if the Purcell factor is increased further and the system reaches the strong coupling regime, the cavity mode and exciton hybridize into two polariton states where the phonon environment will drive incoherent transitions
from the upper polariton state to the lower one, thereby reducing the indistinguishability\cite{Smith2017,Denning:20}. This leads to an inherent trade-off between efficiency and indistinguishability\cite{Smith2017}, since the efficiency cannot be further increased by Purcell enhancement without compromising the indistinguishability. While this trade-off limits the maximum indistinguishable photon generation efficiency of the standard micropillar SPS design to $\sim$ 0.95 \cite{Wang2020_PRB_Biying}, it was recently shown that this limit can be circumvented by suppressing the background emission $\Gamma_{\rm B}$ while maintaining adequate Purcell enhancement, allowing for improved efficiency without compromising the indistinguishability\cite{Gaal2022}. It is thus of interest to increase $\beta$ by decreasing $\Gamma_{\rm B}$ rather than just increasing $F_{\rm p}$.Indeed, several broadband designs, including the photonic nanowire\cite{Bleuse2011, Gregersen2016, Gregersen:10, Gregersen:08, Claudon2010, Claudon13, Munsch2013,Bulgarini2014} and the photonic crystal waveguide\cite{Lecamp2007b, MangaRao2007, Arcari2014,Zhou_2022,Beatrice22,ostfeldt22}, allow for control and suppression of the background emission $\Gamma_{\rm B}$, improving the $\beta$ factor and, in turn, the collection efficiency in a broad spectral range. Additionally, the circular Bragg grating, or "bullseye" cavity\cite{Davanço11,Ates12,Yao2018, Liu2019,Wang2019a, Wang2019b,Rickert:19,Dusanowska20}, and the nanopost\cite{Kotal2021,Jacobsen23} cavity exhibit similar suppression effects. 

\begin{figure}[h!]
	\begin{subfigure}{1\linewidth}
		\centering
  		\includegraphics[width= 0.95 \linewidth]{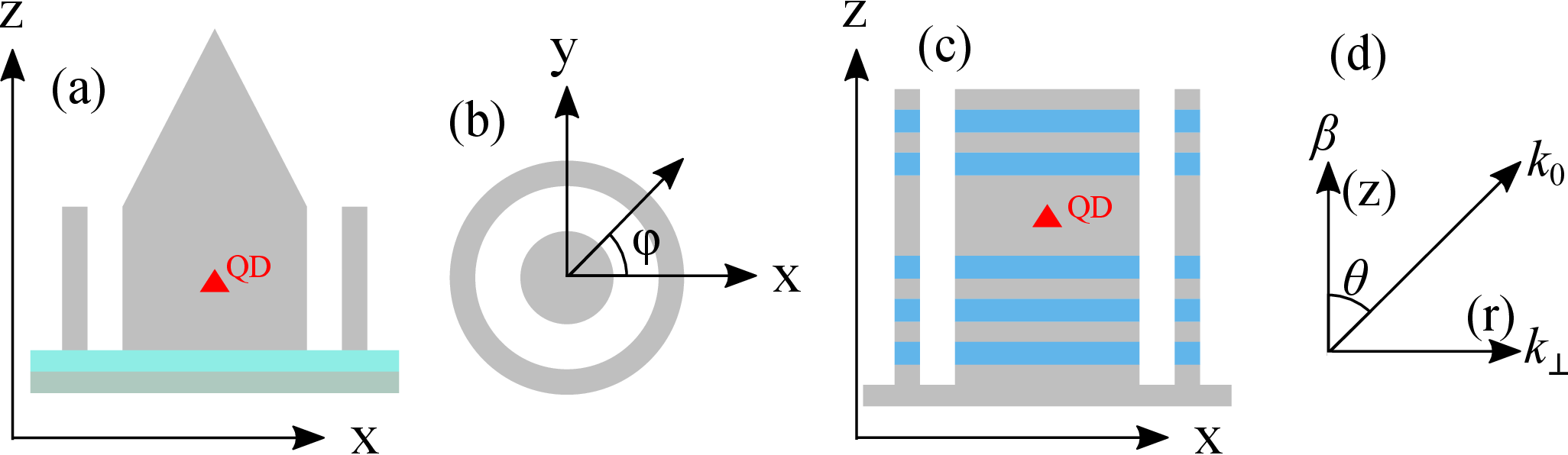}
	\end{subfigure}
	\caption{(a) Sketch of the photonic nanowire SPS with a needle tapering surrounded by 1 ring. (b) Cross sectional view of the nanowire with 1 ring. (c) Sketch of the micropillar SPS surrounded by 1 ring. The cross sectional view of the micropillar is similar to the nanowire. (d) Sketch of the relation between the propagation constant $\beta$ (not to be confused with the $\beta$ factor) and the in-plane $k$ value $k_{\perp}$ for the propagating radiation modes. $\beta$ also defines the angle $\theta$ of propagation with respect to the z axis as $\beta=k_0\cos(\theta)$.}
	\label{intro_figure}
\end{figure}

In this work, we introduce circular Bragg reflectors with 1-3 periods around nanowires or micropillars to further decrease the SE rate $\Gamma_{\rm B}$ into background modes. The circular Bragg reflector consists of alternating layers of air and high-index materials, thus forming concentric rings separated by air gaps. We provide a detailed physical description of how rings around an infinite nanowire can influence the emission into radiation modes. We demonstrate how the few-period circular Bragg reflector around an infinite nanowire can increase the $\beta$ factor of the fundamental $\mathrm{HE}_{11}$ mode ($\beta=\Gamma_{\rm HE_{11}}/\Gamma_{\rm T}$) up to 0.999, thanks to the additional suppression of the emission into radiation modes caused by a photonic bandgap effect. We apply this strategy in the practically relevant case of the finite-sized SPS based on tapered nanowires sketched in Figs.\ (\ref{intro_figure}a-\ref{intro_figure}b). Finally, we also show the beneficial effects of adding optimized rings around micropillars, as shown in Fig.\ (\ref{intro_figure}c). We thus improve upon two well established SPS designs.

\section*{Theory}  \label{Theory}
To calculate the figures of merit, we compute the total and collected power from the source by solving Maxwell’s equations in the frequency domain, modeling the QD as a classical electric point dipole as commonly done in the literature\cite{Schneider:18}. Within the dipole approximation, the emission rate of the QD is related to the classical emitted power of a dipole as $\Gamma/\Gamma_{\rm 0}=P/P_{\rm 0}$\cite{novotny2012principles}, where $P$ is the emitted power of the classical dipole and $P_{\rm 0}$ is the power emitted in a bulk medium. The $\beta$ factor is then defined as $\beta=P_{\rm mode}/P_{\rm T}$, where $P_{\rm mode}$ is the power emitted into our mode of interest (for instance, the cavity mode or the fundamental $\mathrm{HE}_{11}$ mode), and $P_{\rm T}$ is the total emitted power. The collection efficiency is then defined as $\varepsilon_{g}=P_{\mathrm{collected},g}/P_{\rm T}$, where $P_{\mathrm{collected},g}$ is the power collected in the far-field by a lens with specific numerical aperture ($\rm NA$) and taking into account the overlap with a Gaussian shaped field profile\cite{Wang2020_PRB_Biying}. The SMM efficiency is then given by $\varepsilon_{g,\mathrm{SMM}}=\gamma_g\beta$, where $\gamma_g=P_{\mathrm{collected,mode},g}/P_{\rm mode}$. The current distribution of the point dipole is given by $\mathbf{J}(\mathbf{r})=-i\omega\mathbf{p}\delta(\mathbf{r}-\mathbf{r}_{\rm d})$, where $\omega$ is the angular frequency, $\mathbf{r}_{\rm d}$ is the position of the QD, and $\mathbf{p}$ is the dipole moment. We generally assume that the dipole is placed on-axis and the dipole orientation is in-plane; however, we also consider certain cases of off-axis dipoles. 

In our initial analysis, we consider geometries with uniformity along a propagation z axis, and the modeling is performed using a modal expansion method, where the field is expanded on eigenmodes of uniform geometries along z. We use a novel method to analytically construct the radiation modes of a multilayered cylindrical structure inspired by previous work\cite{Yeh:78,Manenkov1970,Snyder1971,SammutPHD,Nyquist1981,Vassallo81,Shevchenko82,Sammut1982,Vassallo83,Snyder1983,Tigelis1987,Morita1988,Alvarez2005}. This method, discussed in detail in Supplementary (S1), allows for computation efficiency and direct insight into the governing physics. The eigenmodes consist of a finite discrete set of modes and an infinite continuum of modes\cite{Nyquist1981}. In the literature, the discrete set of modes is typically referred to as guided modes\cite{Yariv_6th} and the continuum as radiation modes\cite{Nyquist1981}. HE and EH modes are two standard ways of classifying the guided modes originating from fiber theory\cite{Yariv_6th}, and $\mathrm{HE}_{11}$ is the first guided mode, i.e., the fundamental $\mathrm{HE}_{11}$ mode. All structures considered in this paper feature rotational symmetry, and the expression for the electrical field in a layer with uniformity along $z$ is then given by

	\begin{equation}
		\begin{split}
	\textbf{E}^{\pm}(r,\phi,z)&=\sum_{p=1}^2\Biggl(\sum_{n=1}^{N}a^{\pm}_{p,n}\textbf{e}^{\pm}_{p,n}(r,\phi)e^{\pm i\beta_{n} z}+\sum_{m=0}^{\infty}\sum_{s=1}^2\int_0^{\infty}a^{\pm}_{p,m,s}(k_{\perp})\textbf{e}^{\pm}_{p,m,s}(r,\phi,k_{\perp})e^{\pm i\beta(k_{\perp}) z}d k_{\perp}\Biggr),
 \label{Eexpand}
 \end{split}
 \end{equation} where $p$ refers to the two orthogonal polarizations (such as horizontal (H) and vertical (V) polarization), $\pm$ refers to forward or backwards propagation, $\beta_n$ and $\beta(k_{\perp})$ are the propagation constants (not to be confused with the $\beta$ factor), $\textbf{e}(r,\phi)$ is the mode profile, $a$ is the expansion coefficient, $N$ is the total number of guided modes, $m$ is the angular momentum, $k_{\perp}$ is the in-plane $k$ value, and $s$ refers to the two orthogonal solutions for the radiation modes. For each $n$, there is also an associated value of $m$. The guided modes are confined to regions with high refractive index, usually the core of the structure, and decay in regions with low refractive index, such as air regions. Their propagation constant $\beta_{n}$ satisfies the following relation $n_{\rm max}k_0\geq\beta_{n}>n_{\rm air}k_0$, where $k_0=2\pi/\lambda_0$ is the free-space $k$ vector amplitude ($\lambda_0$ being the free-space wavelength), $n_{\rm max}$ is the maximum refractive index of the layer, and we assume that the surrounding background material is air. The propagation constant of the radiation modes is related to the in-plane $k$ value as $\beta(k_{\perp})=\sqrt{(n_{\rm air}k_0)^2-k_{\perp}^2}$. The radiation modes can be viewed as perturbed versions of the modes in a homogeneous air medium (where the two $s$ values would correspond to transverse electric (TE) and transverse magnetic (TM) modes) and are generally not confined to any specific regions. The continuum of radiation modes can be split into two parts: the propagating radiation modes $n_{\rm air}k_0\geq\beta(k_{\perp})>0$ and the evanescent radiation modes $0>\beta(k_{\perp})$. For the propagating radiation modes, the propagation constant also defines the angle of propagation with respect to the z-axis: $\beta=k_0\cos(\theta)$ (with $n_{\rm air}=1$), which is sketched in Fig.\ (\ref{intro_figure}d). The expansion coefficients for the field generated by a dipole can be derived from the Lorentz reciprocity theorem\cite{Nyquist1981,NUMERICALMETHOD2014} as discussed in Supplementary (S1.5). The emitted power can then be evaluated as\cite{novotny2012principles}
\begin{equation}
	\begin{split}
		P = - \frac{1}{2}\int_V{\rm Re}[{{\bf J}^*({\bf r}_{\rm})\cdot{\bf E}(\bf r_{\rm})}]dV= \frac{\omega}{2} {\rm Im}[{{\bf p} \cdot{\bf E}(\bf r_{\rm d})}] .    
	\end{split}
	\label{Power}
\end{equation} Standard semiconductor InGaAs QDs feature an in-plane dipole moment\cite{Heindel2023}, and when the QD is placed on-axis, the in-plane oriented dipole only couples to eigenmodes with angular momentum of $m=1$, since the in-plane electrical field components of eigenmodes with $m\neq1$ vanish at $r=0$. Additionally, the resulting power is identical for the two in-plane dipole orientation due to the rotational symmetry, and thus only one polarization $p$ needs to be considered. Still, the polarization of the emitted photons can be controlled, for instance, by selectively exciting one of the two orthogonal dipole states of a neutral exciton in a QD, each corresponding to a specific linear polarization of the emitted photon (H or V). This state preparation can be achieved with advanced state preparation schemes, such as LA phonon assisted\cite{Thomas2021}, SUPER\cite{Karli2022,Piccinini2025}, or stimulated TPE\cite{Sbresny2022,Wei2022,Yan2022}. These three schemes can prepare the emitter in a specific linear dipole (H or V) and require laser pulses that are spectrally separated from the QD emission line, making it possible to reject the excitation laser with spectral filtering.

In the infinite structures, the evanescent modes do not contribute to the power, and thus one can replace the integral over all $k_{\perp}$ in Eq.\ (\ref{Eexpand}) with $\int_0^{k_0}$. Assuming on-axis dipole placement, the total emitted power in an infinite structure can then be written as 
\begin{equation}
	\begin{split}
		P_{\rm T} = \sum_{n=1}^NP_{g,n}+\sum_{s=1}^2\int_0^{k_0}p_s(k_{\perp})d k_{\perp}=P_{\rm G}+P_{\rm R} = P_{\rm HE_{11}} + P_{\rm HG} +P_{\rm R} ,
	\end{split}
	\label{Powertot}
\end{equation} where $P_{\rm G} = P_{\rm HE_{11}} + P_{\rm HG}$ is the power into all guided modes, $P_{\rm HG}$ is the power into any higher-order guided modes excluding the HE$_{11}$ mode, and $P_{\rm R}$ is the power into the radiation modes. In the single-mode regime relevant for the photonic nanowire, the power into background modes is $P_{\rm B} = P_{\rm R}$, whereas in the large-diameter regime relevant for the micropillar the power $P_{\rm B} = P_{\rm HG} + P_{\rm R}$ includes contributions from both the higher-order guided modes and the radiation modes. We choose an orthogonalization $s$ such that $p_{s=1}(k_{\perp})=0$, and we will often simply use $p(k_{\perp})=p_1(k_{\perp})+p_2(k_{\perp})$, where $p(k_{\perp})$ is the power spectrum for the radiation modes. In the remainder of the paper, we assume that the dipole is placed on-axis unless stated otherwise.

The modal expansion method can also handle the finite-sized structures under study in Sections "\hyperref[Photonic nanowire]{Finite-size photonic nanowire}" and "\hyperref[Micropillar]{Micropillar}". In the Fourier Modal Method, the geometry is split into layers with uniformity along the z axis and the fields on each side of a layer interface are connected using a scattering matrix formalism as discussed in Supplementary (S1). 

In Eq.\ (\ref{betacav}), any contributions from intrinsic non-radiative recombination are omitted. Instead, non-radiative recombination effects are accounted for in the quantum efficiency defined as $\eta_{\mathrm{QE}}=\Gamma_{\rm T}/(\Gamma_{\rm T}+\Gamma_{\rm NR})$, where $\Gamma_{\rm NR}$ is the non-radiative recombination rate. Effectively, the figures or merit should then be corrected by multiplying by $\eta_{\mathrm{QE}}$ and thus written as $\beta_{\rm corrected}=\eta_{\mathrm{QE}}\beta$, $\varepsilon_{g,\rm SMM,corrected}=\gamma_g\beta_{\rm corrected}$, and $\varepsilon_{g,\rm corrected}=\eta_{\mathrm{QE}}\varepsilon_g$. In this paper, we focus on the optics and the radiative rates, not on the intrinsic properties of the emitter. Still, we will address the scenarios where $\eta_{\mathrm{QE}}$ becomes relevant within the scope of the paper.

While the $\beta$ factor and $\gamma_g$ are useful tools for optimization and provide an intuitive understanding of the collection efficiency via the SMM, they are not straight forward to measure in an experiment. Nevertheless, the $\beta$ factor may be estimated experimentally\cite{Thyrrestrup2010}, and $\gamma_{g}$ can then be estimated having also measured $\varepsilon_{g}$.


\section*{Suppression of the spontaneous emission into radiation modes using a circular Bragg reflector} \label{Physics}
Our strategy for suppressing the emission into radiation modes is to implement a photonic bandgap in the radial direction by introducing concentric rings. Beyond SPSs, multilayered cylindrical structures have been explored for applications such as lasers\cite{Scheuerlasing,Scheuerlaser2,Scheuerlasing3,Scheuer:07,Jebali:07,Scheuer10}, fibers\cite{Yeh:78,Doran1983,Johnson:01,braggfiber2}, and sensing\cite{Scheuersensing}, and their underlying physics has been described in various studies\cite{Jiang93,Jiang94,jiang1994cylindrical,Nikolaev1999,Nikolaev1999v2,Kaliteevski00,Kaliteevskii2000,Ochoa00,Scheuer03v1,Scheuer03v2}. Just as plane waves are reflected at an interface between two materials, so are cylindrical waves. However, a key distinction is that cylindrically propagating waves, described by the Bessel functions of the first and second kind, $J_m(k_{\perp}r)$ and $Y_m(k_{\perp}r)$, are non-periodic. Consequently, the standard quarter-wavelength ($\lambda_{\rm eff}/4$) layer thickness condition used in planar photonic bandgap structures becomes inaccurate in cylindrical systems and should be replaced with a condition taking into account the initial non-periodic evolution of the Bessel functions. In this way, the reflection may be maximized for a specific $k_{\perp}$ value leading to a photonic bandgap effect with suppression of the SE into this particular $k_{\perp}$ mode. 

To suppress the full continuum $k_{\perp}\in[0,k_0]$ of cylindrically propagating waves, we need to consider all values of $k_{\perp}$ in this interval. Fortunately, this continuum transforms into a much narrower interval in the semiconductor material. Indeed, for a material with refractive index $n_{\rm mat}$, the in-plane material $k$ value $k_{\perp,\rm mat}$ depends on the in-plane free-space $k$ value $k_{\perp}$ as $k_{\perp,\rm mat}=\left({k_{\perp}^2+(n_{\rm mat}^2-1)k_0^2}\right)^{1/2}$, such that the relevant interval to be considered becomes $k_{\perp,\rm mat}\in[{(n_{\rm mat}^2-1)}^{1/2} k_0,n_{\rm mat}k_0]$. We observe that a high refractive index contrast not only increases the reflection and the spectral width of the photonic bandgap but also reduces the $k_{\perp,\rm mat}$ interval for which the SE should be suppressed.

\begin{figure}[h!]
	\begin{subfigure}{1\linewidth}
		\centering
  		\includegraphics[width= 0.80 \linewidth]{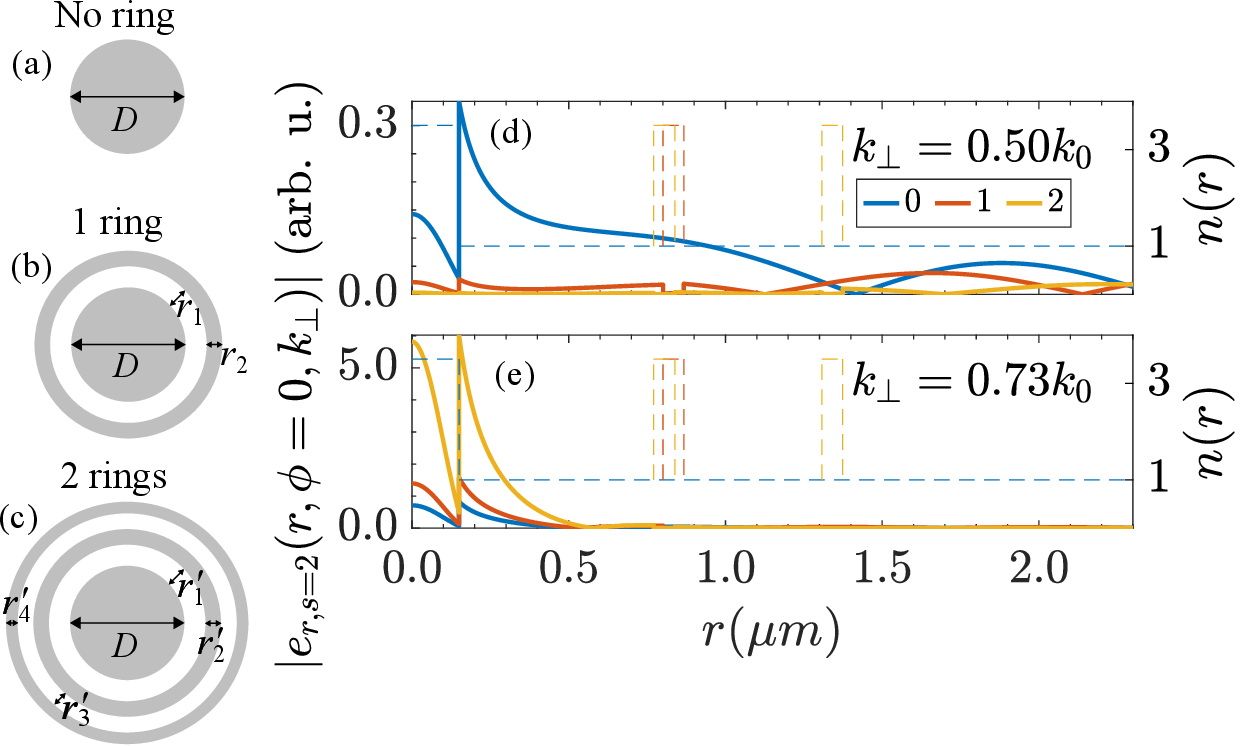} 
	\end{subfigure}
	\caption{Cross-sectional view of the nanowire with 0 (a), 1 (b), and 2 rings (c). The absolute value of the radial electrical field component (full lines) for the radiation mode with $k_{\perp}=0.5k_0$ (d) and $k_{\perp}=0.73k_0$ (e) is shown as a function of the radius, $r$, for 0, 1, and 2 ring(s) with $D=\SI{300}{nm}$. The dashed lines indicate the refractive index profile for the three configurations.}
	\label{field_radiationv2}
\end{figure}

Let us now consider an infinitely long GaAs nanowire with a diameter of $D=\SI{300}{nm}$ and compare the performance for the configurations with 0, 1, and 2 rings shown in Figs.\ (\ref{field_radiationv2}a-\ref{field_radiationv2}c). 
We use a wavelength of $\lambda_0=\SI{895}{nm}$  and a refractive index of GaAs of $n_{\rm GaAs}=3.5015$ throughout the paper. We then select the radiation mode with $k_{\perp}=0.5k_0$ and optimize the ring thicknesses to minimize $p(k_{\perp}=0.5k_0)$. The fast computation speed of the analytical method allows us to simply scan the entire geometrical parameter space to find the minimum of $p(k_{\perp}=0.5k_0)$.  

 The influence of the rings on the field profile for the radiation mode $k_{\perp}=0.5k_0$ is presented in Fig.\ (\ref{field_radiationv2}d). The electric mode profile is strongly suppressed at $r=0$ for 1 ring and even more for 2 rings. The optimized ring parameters for 1 ring are $r_1=\SI{650.7}{nm}$ and $r_2=\SI{66.6}{nm}$, and for 2 rings $r'_1=\SI{621.1}{nm}$, $r'_2=\SI{67.2}{nm}$, $r'_3=\SI{469.4}{nm}$, and $r'_4=\SI{66.1}{nm}$. This also shows that adding an extra ring can change the optimal dimensions of the first ring. However, minimizing $p(k_{\perp}=0.5k_0)$ does not ensure that $p(k_{\perp})$ is reduced for all values of $k_{\perp}$ and in Fig.\ (\ref{field_radiationv2}e), the field is shown for the radiation mode with $k_{\perp}=0.73k_0$. Here, the field is actually enhanced at $r=0$, and thus $p(k_{\perp}=0.73k_0)$ increases. This shows the importance of considering the entire continuum and not just a selected value of $k_{\perp}$.

The $\beta$ factor, given by $\beta=P_{\rm HE_{11}}/(P_{\rm R}+P_{\rm G})$ in the infinite nanowire case, depends not only on $P_{\rm R}$ but also on the power emitted into the fundamental $\mathrm{HE}_{11}$ mode $P_{\rm HE_{11}}$ and into all guided modes $P_{\rm G}$. While we aim to use the rings to suppress $P_{\rm R}$, the rings can in fact also influence the guided modes and thus $P_{\rm G}$. However, the physics of the guided modes differs from the radiation modes as they are not propagating waves in the radial direction. In the low-index air regions, they are instead described by the modified Bessel functions of $K_{m}(k_{\perp,\rm g}r)$ and $I_{m}(k_{\perp,\rm g}r)$, where $k_{\perp,\rm g}$ is the in-plane $k$ value of the guided mode  (in the outermost air region, only $K_{m}(k_{\perp,\rm g}r)$). If the ring is close enough to the central wire and reaches the evanescent tails of the guided modes, the mode profiles will be altered and the emission rates will be influenced. 
\begin{figure}[h!]
	\begin{subfigure}{1\linewidth}
		\centering
  		\includegraphics[width= 0.65 \linewidth]{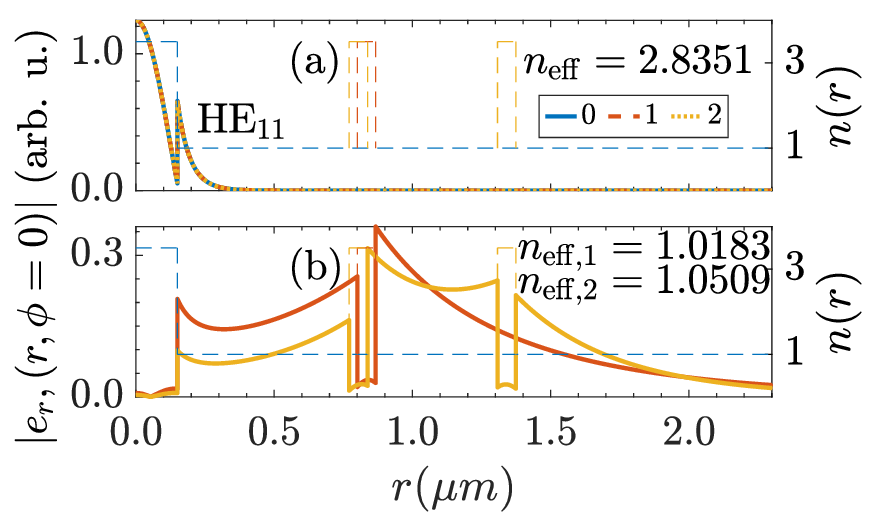} 
	\end{subfigure}
	\caption{(a) The absolute value of the radial electrical field component for the fundamental $\mathrm{HE}_{11}$ mode is shown as a function of the radius, $r$, for 0 (full line), 1 (dashed line), and 2 (dotted line) rings with $D=\SI{300}{nm}$. (b) The field for two selected guided modes for 1 and 2 rings, respectively. The refractive index profile is indicated with the dashed lines as in Figs.\ (\ref{field_radiationv2}d-\ref{field_radiationv2}e).}
	\label{field_guided}
\end{figure} In Fig.\ (\ref{field_guided}a), the field profile for the fundamental $\mathrm{HE}_{11}$ mode is now shown with the exact same parameters obtained from the minimization of $p(k_{\perp}=0.5k_0)$. Here, the ring(s) are far away from the central wire where the mode is confined, and the field profile of the fundamental $\mathrm{HE}_{11}$ mode is thus not influenced and $P_{\rm HE_{11}}$ remains unchanged. However, the ring(s) can, in fact, also introduce additional guided modes despite the bare nanowire only supporting a single guided mode. For the structure with 1 ring, two additional guided modes appear with $n_{\mathrm{eff}}=2.2103$ and $n_{\mathrm{eff}}=1.0183$, while the structure with 2 rings introduces three additional guided modes with $n_{\mathrm{eff}}=2.2201$, $n_{\mathrm{eff}}=2.2094$, and $n_{\mathrm{eff}}=1.0509$. In Fig.\ (\ref{field_guided}b), the field is shown for two of these selected guided modes for 1 and 2 rings, respectively. A small part of the field of these new guided modes extends to the wire axis, thus influencing the SE rate $P_{\rm G}$ into the guided modes. For this reason, we will in the following use $\beta$ as the objective function to be maximized instead of simply minimizing $P_{\rm R}$.


\section*{Infinitely long nanowire} \label{Nanowire}
\begin{figure}[h!]
	\begin{subfigure}{1\linewidth}
		\centering
  		\includegraphics[width= 0.95 \linewidth]{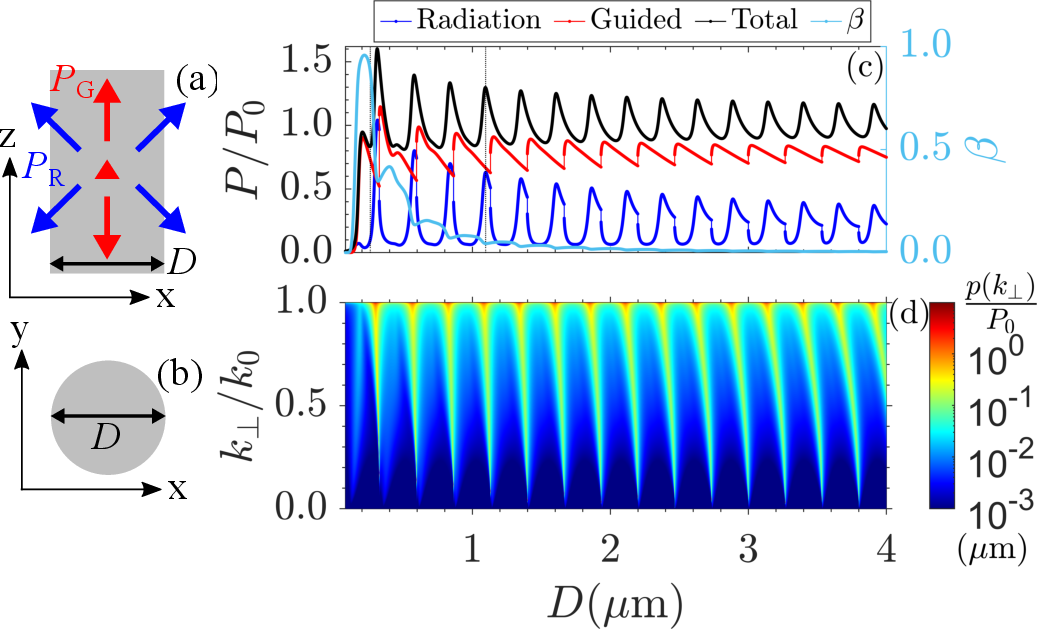} 
	\end{subfigure}
	\caption{(a) Sketch of the infinite nanowire with diameter $D$ with a QD (red triangle) in the center. (b) Cross sectional view of the infinite nanowire. (c) $P/P_0$ as a function of $D$ for the radiation modes, the guided modes, and their sum. The $\beta$ factor (right y-axis) is also shown as a function of $D$. The vertical black dotted lines correspond to the diameters $D_1=\SI{0.26}{\mu m}$ (used in Fig.\ (\ref{1ring_scan})) and $D_2=\SI{1.096}{\mu m}$ (used in Supplementary (S2)). (d) The power spectrum of the radiation modes, $p(k_{\perp})/P_0$, as a function of $D$ and $k_{\perp}/k_0$.}
	\label{nanowire_power_part1}
\end{figure}

To prepare the subsequent discussion of the effect of the rings, we initially discuss the physics of light emission from a QD placed on axis in the infinitely long bare nanowire with diameter $D$ illustrated in Figs.\ (\ref{nanowire_power_part1}a,b).
 The emission rates into the guided modes and the radiation modes, the total emission rate, and the $\beta$ factor are shown in Fig.\ (\ref{nanowire_power_part1}c) as a function of the diameter. For $D \rightarrow 0$, the emission into the radiation modes is suppressed due to the dielectric screening effect\cite{Bleuse2011,Claudon13}. For increasing $D$, the power into the fundamental $\mathrm{HE}_{11}$ mode initially increases as the confinement of the mode increases, while the emission into the radiation modes remains suppressed due to the dielectric screening effect. Here, $\beta$ reaches its maximum value of $\beta_{\rm HE_{11}}=0.9581$ at $D=\SI{0.218}{\mu m}$. As $D$  increases further, the emission into the fundamental $\mathrm{HE}_{11}$ mode decreases due to its increasing mode area, resulting in a decrease in the $\beta$ factor. Additional guided modes emerge in pairs: $\mathrm{EH}_{1,n}$ modes cause discontinuous jumps in the guided SE rate, peaking sharply before declining, whereas $\mathrm{HE}_{1,n+1}$ modes increase more gradually before decreasing \cite{wang2021}. The emission into radiation modes exhibits a semi-periodic pattern of peaks and valleys, with peaks broadening and decreasing in intensity as the diameter increases. These peaks also correspond to peaks in the total emission rate, and at these peaks there will be dips in the $\beta$ factor. Furthermore, discontinuous drops in the radiation SE rate occur precisely when new guided modes appear, ensuring continuity in the total emission rate and showcasing the transition between radiation modes and guided modes.

 To understand the emission into radiation modes, we now examine the power spectrum of the radiation modes $p(k_{\perp})/P_0$ shown in Fig.\ (\ref{nanowire_power_part1}d). The power spectrum describes the emission behavior of each individual radiation mode of $k_{\perp}$ as a function of $D$. At the nanowire-air interface, each radiation mode experiences a reflection, leading to constructive or destructive interference at the dipole position, thereby modulating the emission. As such, each radiation mode has its own semi-periodic emission pattern with peaks and valleys as a function of $D$ based on $k_{\perp}$. Due to the high refractive index, we observe in Fig.\ (\ref{nanowire_power_part1}d) that for smaller diameters, the peaks of individual radiation modes almost coincide at the same value of $D$ across all $k_{\perp}$. This leads to high and narrow peaks in the emission into radiation modes observed in Fig.\ (\ref{nanowire_power_part1}c). However, for increasing diameters, the peaks shift apart in $D$ due to the different periodicities based on $k_{\perp}$. This, in turn, leads to the broadening of the peaks in Fig.\ (\ref{nanowire_power_part1}c).

\section*{Rings} \label{Rings}

\begin{figure}[h!]
	\begin{subfigure}{1\linewidth}
		\centering
		\includegraphics[width= 0.85 \linewidth]{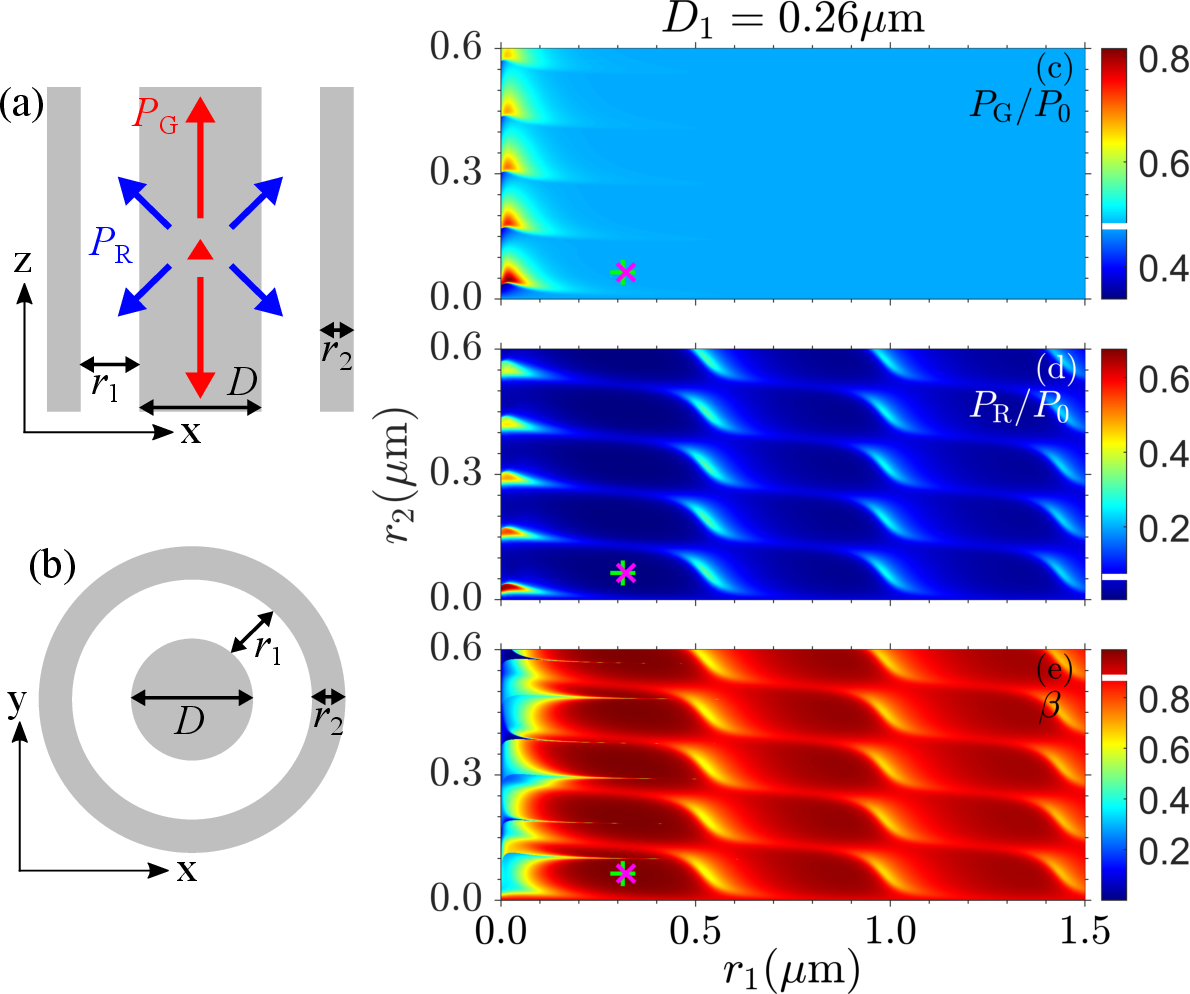} 
	\end{subfigure}
	\caption{(a) Sketch of the infinite nanowire with a QD (red triangle) in the center with one ring and (b) Cross-sectional view. The center nanowire has diameter $D$, the air gap is $r_1$, and the ring thickness is $r_2$. (c-e) $P_{\rm G}/P_0$, $P_{\rm R}/P_0$, and $\beta$ as a function of the air gap $r_1$ and ring thickness $r_2$ for the selected diameter of $D_1=\SI{0.26}{\mu m}$. Moving along the vertical axis for $r_1=\SI{0}{\mu m}$ simply corresponds to increasing the diameter of the infinite nanowire. The white line in the color bars indicates the value of the wire without any rings. The green plus signs correspond to the minimum of $P_{\rm R}/P_0$, while the magenta crosses correspond to the maximum of $\beta$. Discontinuous lines appear in $\beta$ for $r_1<\SI{0.5}{\mu m}$. These discontinuities are not due to discontinuities in $P_\mathrm{T}$, instead the definition of the fundamental $\mathrm{HE}_{11}$ mode becomes blurry in the cases of a small diameter in the proximity of a thick ring, making it difficult to select the correct mode.}
	\label{1ring_scan}
\end{figure}

We now investigate the effect of adding initially a single ring around the nanowire, as shown in Figs.\ (\ref{1ring_scan}a-\ref{1ring_scan}b). 
The normalized emission into the guided modes $P_{\rm G}/P_0$, into radiation modes $P_{\rm R}/P_0$, and the $\beta$ factor are shown in Figs.\ (\ref{1ring_scan}c-\ref{1ring_scan}e) as a function of $r_1$ and $r_2$ for a selected fixed diameter of $D_1=\SI{0.26}{\mu m}$ marked by the first vertical dotted line in Fig.\ (\ref{nanowire_power_part1}c). For the guided modes in Fig.\ (\ref{1ring_scan}c), it can be observed that the ring must be close to the center wire and the evanescent tail of the guided modes to influence the emission rate. The extending tails of slightly larger $r_1$ values seen in Fig.\ (\ref{1ring_scan}c) are resulting from weakly confined higher-order modes with a small propagation constant. These modes decay the slowest outside the center wire and are thus perturbed by a ring placed further away. As discussed earlier, weakly confined ring modes extending to the wire axis can also influence the emission rate. 

On the other hand, the emission rate $P_{\rm R}/P_0$ for the radiation modes presented in Fig.\ (\ref{1ring_scan}d) is affected across the entire parameter space. It displays a semi-periodic interference pattern as a function of $r_1$ and $r_2$, with shorter periodicity in $r_2$ due to the large refractive index of GaAs. The strongest effects are observed within the first period of varying $r_1$ and $r_2$, and here the maximum in $\beta$, indicated by the magenta crosses, and the minimum in $P_{\rm R}/P_0$, indicated by the green plus signs, are found. The magenta cross and the green plus overlap almost perfectly, demonstrating that in this case, $D_1=\SI{0.26}{\mu m}$, the $\beta$ factor improvement from $\beta=0.8785$ to $\beta_{\mathrm{1 ring}}=0.9890$ directly results from inhibition of SE into radiation modes. Thus, the effect of the photonic bandgap can already be observed for one ring. We present a similar scan in Supplementary (S2) for a larger diameter of $D_2=\SI{1.096}{\mu m}$ corresponding to a peak in $P_{\rm R}/P_0$ indicated by the second vertical dotted line in Fig.\ (\ref{nanowire_power_part1}c).

\subsection*{$\beta$ factor optimization}

We will now optimize the $\beta$ factor (of the fundamental $\mathrm{HE}_{11}$ mode) as a function of the center diameter with respect to the ring dimensions. For a single-ring configuration, the fast computation speed of the analytical method allows us to easily cover the entire $(r_1,r_2)$ parameter space.
\begin{figure}[h!]
	\begin{subfigure}{1\linewidth}
		\centering
		\includegraphics[width= 1 \linewidth]{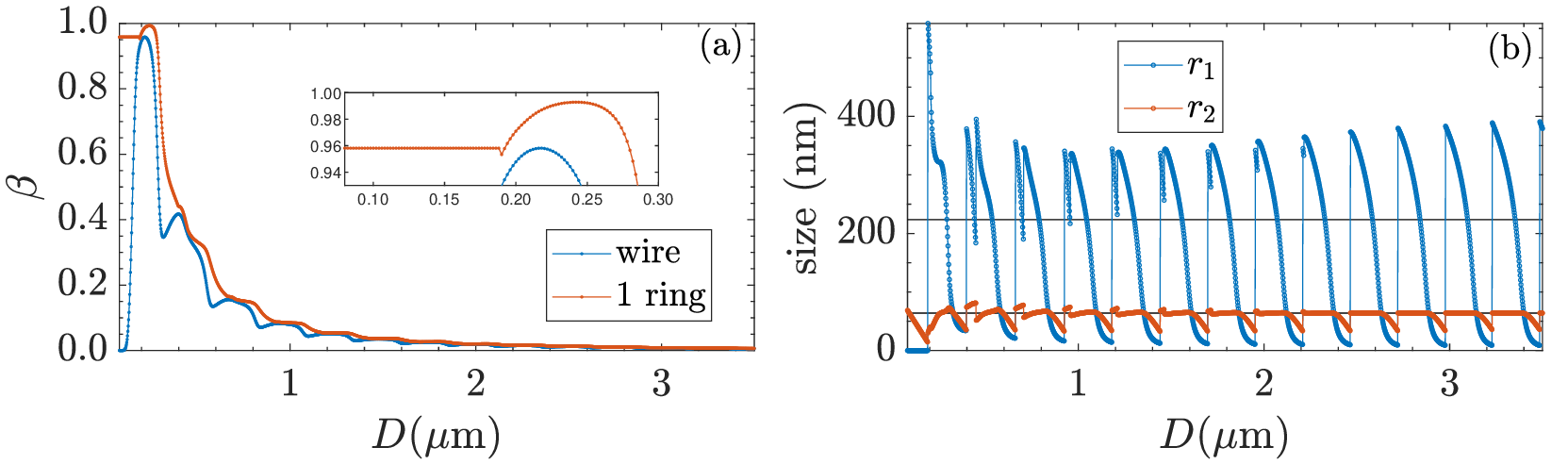} 
	\end{subfigure}
	\caption{(a) $\beta$ factor as a function of the diameter for the wire alone and the wire with one optimized ring. (b) Optimized air gap $r_1$ and ring thickness $r_2$ as a function of the diameter. The two horizontal black lines indicate $r_1=\lambda/(4n_{\mathrm{air}})$ and $r_2=\lambda/(4n_{\mathrm{GaAs}})$, i.e., standard quarter-wave thicknesses.}
	\label{beta_opt_1ring}
\end{figure} Fig.\ (\ref{beta_opt_1ring}a) shows the $\beta$ factor as a function of $D$ for both the bare wire and the wire with one optimized ring. The optimized ring improves the $\beta$ factor for all diameters, with a significant increase in the maximum $\beta$ factor from $\beta_{\rm wire}=0.9581$ at $D=\SI{0.218}{\mu m}$ to $\beta_{\rm 1 ring}=0.9928$ at $D=\SI{0.242}{\mu m}$ with $r_1=\SI{323.4}{nm}$ and $r_2=\SI{59.0}{nm}$. This reduces losses to other modes from $1-0.9581=0.0419$ to $1-0.9928=0.0072$, a factor of 5.8 improvement. For larger diameters, $\beta$ decreases due to increased emission into higher-order guided modes, which the ring does not control, and due to reduced emission into the fundamental $\mathrm{HE}_{11}$ mode. Nevertheless, we will later demonstrate the ring's beneficial effect on the performance of the large-diameter QD-micropillar SPS. The corresponding optimized parameters for the air gap $r_1$ and ring thickness $r_2$ are shown in Fig.\ (\ref{beta_opt_1ring}b) as a function of $D$. For $D<\SI{0.218}{\mu m}$, the optimization effectively increased the diameter to the fixed value of $D=\SI{0.218}{\mu m}$ by setting $r_1=0$, thus keeping the $\beta$ factor constant. Generally, the optimized parameters follow a semi-periodic pattern, with $r_1$ exhibiting larger variations than $r_2$ due to the much lower refractive index of air compared to GaAs. Importantly, Fig.\ (\ref{beta_opt_1ring}b) shows that using standard quarter-wave thicknesses, $r_1=\lambda/(4n_{\mathrm{air}})$ and $r_2=\lambda/(4n_{\mathrm{GaAs}})$, indicated by the two horizontal black lines, generally does not provide the optimal suppression of $P_{\rm R}$ due to the non-periodic behavior of the Bessel functions and the continuum in $k_{\perp}$ as discussed earlier. The detailed behavior of $P_{\rm R}/P_0$, $P_{\rm G}/P_0$, $P_{\rm T}/P_0$, and $P_{\rm HE_{11}}/P_0$ as a function of $D$ is provided in Supplementary (S2) for the two configurations.

We proceed to optimize the $\beta$ factor for configurations with two and three rings in the small-diameter regime. For each configuration, we perform two types of optimization: (i) a fully flexible optimization where all air gaps and ring thicknesses are treated as free parameters, and (ii) a constrained optimization where all air gaps and ring thicknesses are respectively identical (referred to as the "fixed" case). The fully flexible approach involves 2N free parameters for a given diameter (N being the number of rings), while the fixed case involves only 2 parameters. Details on the numerical optimization are given in Supplementary (S3\&S4). Although ideal optimization of the $\beta$ factor requires that all parameters are free, we also consider the less numerically demanding fixed case, commonly used in SPS designs featuring concentric rings such as the "bullseye" SPS \cite{Rickert:19}.

\begin{figure}[h!]
	\begin{subfigure}{1\linewidth}
		\centering
		\includegraphics[width= 1 \linewidth]{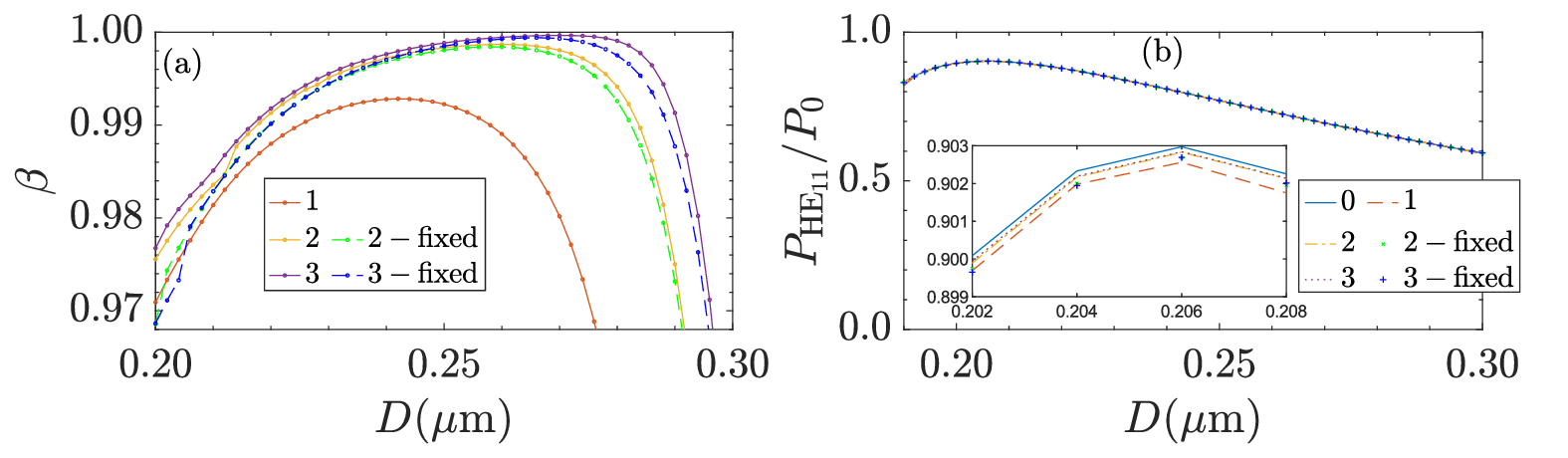} 
	\end{subfigure}
	\caption{(a) Optimized $\beta$ factor and (b) $P_{\mathrm{HE}_{11}}/P_0$ as a function of the wire diameter for 1-3 rings with free parameters and for 2-3 rings with with fixed parameters. For $P_{\mathrm{HE}_{11}}/P_0$, we also included the case of no (0) rings.} 
	\label{beta_opt_2_3_combined}
\end{figure} The $\beta$ factor as a function of $D$, for both free and fixed parameters for configurations with 1 to 3 rings, is presented in Fig.\ (\ref{beta_opt_2_3_combined}a). The $\beta$ factor is increased above 0.99 over a wide diameter range and even exceeding 0.999 when the number of rings is increased. Leaving all parameters to be free slightly increases the maximum $\beta$ factor and slightly expands the high-$\beta$ range. Still, even though the radiation modes are cylindrical waves and form a continuum, the penalty of fixing the parameters is fairly small. We observe that the emission into the fundamental HE$_{11}$ mode $P_{\mathrm{HE}_{11}}/P_0$ presented in Fig.\ (\ref{beta_opt_2_3_combined}b), is minimally affected due to the lack of influence of the rings on the mode profile, similar to the case shown 
in Fig.\ (\ref{field_guided}a). At the optimal $\beta$ diameter of the bare wire ($D=\SI{0.218}{\mu m}$) or at the maximum of $P_{\mathrm{HE}_{11}}/P_0$ ($D=\SI{0.206}{\mu m}$), the fully optimized configuration with 3 rings achieves $\beta=0.9907$ or $\beta=0.9824$, respectively. The ability of the optimized rings to effectively suppress the power spectrum of the radiation modes, $p(k_{\perp})/P_0$, across all $k_{\perp}$ values is presented in detail in Supplementary (S5). 

\begin{table*}[ht]
\centering
\begin{tabular}{|l|l|l|l|l|l|l|}
\hline
                              & no rings  & 1 ring          & 2 rings         & 2 rings - fixed & 3 rings         & 3 rings -  fixed \\ \hline
$\mathrm{max}(\beta)$             & $0.9581$ & $0.9928$        & $0.9987$        & $0.9984$                      & $0.9997$        & $0.9994$                      \\ \hline
$P_{\mathrm{HE}_{11}}/P_0$ & $0.8845$ & $0.8010$        & $0.7317$        & $0.7393$                      & $0.6945$        & $0.7092$                      \\ \hline
$P_{\mathrm{R}}/P_0$       & $0.0387$ & $0.0051$        & $6.7\text{e-}4$ & $7.1\text{e-}4$               & $1.3\text{e-}4$ & $1.8\text{e-}4$               \\ \hline
$P_{\mathrm{HG}}/P_0$      & 0        & $6.6\text{e-}4$ & $2.8\text{e-}4$ & $4.5\text{e-}4$               & $1.0\text{e-}4$ & $2.3\text{e-}4$               \\ \hline
$D (\mathrm{\mu m})$          & $0.218$  & $0.242$         & $0.260$         & $0.258$                       & $0.270$         & $0.266$                       \\ \hline
$r_1 (\mathrm{nm})$           &          & $323$        & $331$        & $318$                      & $359$        & $333$                      \\ \hline
$r_2 (\mathrm{nm})$           &          & $59$         & $62$         & $62$                       & $64$         & $64$                       \\ \hline
$r_3 (\mathrm{nm})$           &          &                 & $245$        & $318$                      & $246$        & $333$                      \\ \hline
$r_4 (\mathrm{nm})$           &          &                 & $67$         & $62$                       & $72$         & $64$                       \\ \hline
$r_5 (\mathrm{nm})$           &          &                 &                 &                               & $234$        & $333$                      \\ \hline
$r_6 (\mathrm{nm})$           &          &                 &                 &                               & $67$         & $64$                       \\ \hline
\end{tabular}
\caption{Maximum $\beta$ factor and corresponding values of $P_{\mathrm{HE}_{11}}/P_0$, $P_{\mathrm{R}}/P_0$, $P_{\mathrm{HG}}/P_0$, diameter $D$, and ring dimensions for the optimized structures. While the numerical optimization provides additional digits for the optimal parameters, the output values were not affected by restricting the precision to 1 nanometer.}
 	\label{table1}
\end{table*} We present the optimized $\beta$ factor, $P_{\rm HE_{11}}/P_0$, $P_{\rm R}/P_0$, $P_{\rm HG}/P_0$, and the corresponding diameters and ring dimensions in Table \ref{table1}. $P_{\rm HE_{11}}/P_0$ decreases with additional rings due to the increase in the optimized wire diameter observed in Fig.\ (\ref{beta_opt_2_3_combined}). However, this reduction is compensated for by the suppression of $P_{\rm R}$. On the other hand, in the presence of non-radiative recombination effects (see Supplementary (S6)), the quantum efficiency $\eta_{\rm QE}$ will be affected, and the optimal diameter that maximizes, in this case, the product $\eta_{\rm QE}\beta$ shifts to lower values where $P_{\rm HE_{11}}$ is higher. Despite this shift, the rings continue to enhance $\eta_{\rm QE}\beta$ for all diameters. Notably, the rings have reduced $P_{\rm R}/P_0$ to such a degree that the contribution $P_{\rm HG}/P_0$, due to higher-order guided modes introduced by the rings themselves as shown in Fig.\ (\ref{field_guided}b), becomes comparable in magnitude.

So far, we have considered the performance of perfect geometries without fabrication imperfections. However, in Supplementary (S7), we show that for deviations in the ring thicknesses and central mesa radius of up to 5 nm, the maximum $\beta$ factor for 3 rings can be maintained above 0.9981, demonstrating reasonable robustness towards mesa and ring thickness variations. Additionally, the QD position in a real device may not be aligned with the center axis, in particular for top-down nanofabrication from randomly positioned epitaxial QDs\cite{Fons2018}. When the QD is placed off-axis, the $\beta$ factor will generally decrease, and now the dipole orientation also matters as the rotational symmetry breaks. We present simulations for off-axis dipoles for the optimized inner mesa and ring parameters provided in Table \ref{table1} in Supplementary (S8). For a radially oriented dipole and an offset of 50 nm (75 nm), we obtain a $\beta$ of 0.9493 (0.9242) for the bare nanowire, a $\beta$ of 0.9770 (0.9454) for 1 ring, a $\beta$ of 0.9717 (0.9294) for 2 rings, and a $\beta$ of 0.9644 (0.9162) for 3 rings. We note that for the 75 nm offset, the configurations with 2 (3) rings do not provide improved performance compared to the bare wire due to the lower value of $P_{\rm HE_{11}}$ for the optimized inner mesa diameter of $D$ = 260 nm ($D$ = 270 nm). The tolerance for a large offset can be improved by choosing a reduced central mesa diameter with a higher $P_{\rm HE_{11}}$. The configuration with 3 rings optimized at $D=\SI{0.218}{\mu m}$ results in a $\beta$ of 0.9862 (0.9702) for an offset of 50 nm (75 nm). For an azimuthally oriented dipole, the performance is significantly worse for all configurations, $\beta\sim0.5$, due to a significant contribution from the $\mathrm{TE}_{01}$ guided mode\cite{Claudon13}. This problem can be circumvented by selectively exciting the linear dipole transition along the radial axis of a neutral exciton as described in the Theory section. Alternatively, bottom-up growth approaches typically offer deterministic placement of the QD on the center axis\cite{Leandro2018,Choi2021}.

To summarize, the main result of this section is that the $\beta$ factor has improved from 0.9581 without rings to 0.9997 for the design with 3 rings. The geometrical parameters of our optimal 3-ring design are provided in Table \ref{table1}.

\section*{Finite-size photonic nanowire} \label{Photonic nanowire}
\begin{figure}[h!]
	\begin{subfigure}{1\linewidth}
		\centering
		\includegraphics[width= 0.95 \linewidth]{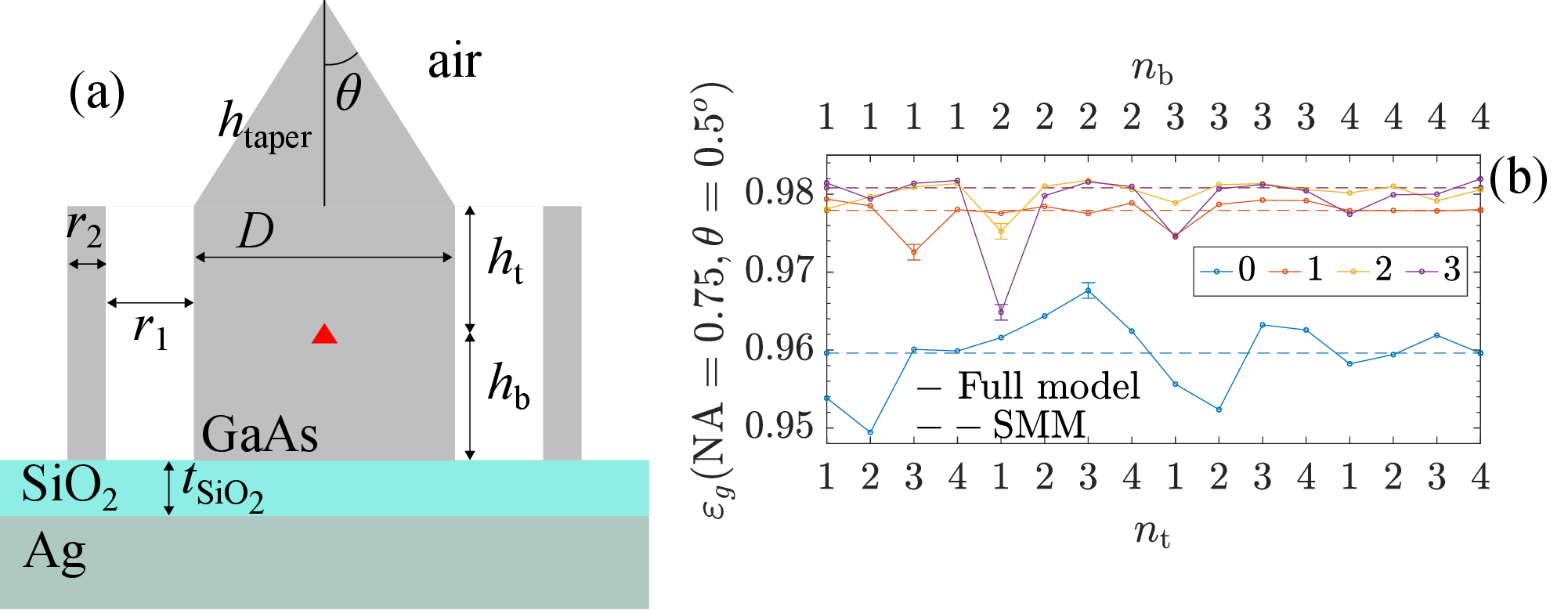}
	\end{subfigure}
	\caption{(a) Sketch of the photonic needle SPS with ring(s) and its geometrical parameters. (b) Collection efficiency $\varepsilon_g (\mathrm{NA}=0.75)$ as a function of the 16 combinations of $n_{\rm t}$ and $n_{\rm b}$ for 0-3 rings computed for a tapering angle of $\theta=0.5^{\circ}$. The dotted line is calculated using the SMM while the solid line is the full model. The predicted SMM values for 2 and 3 rings are almost identical and thus the lines are overlapping. The estimated numerical error is 0.0010 for all data points including the SMM which is showcased by the four error bars.}
	\label{needle_eps}
\end{figure}

We now apply the concept of suppression of background emission using rings, demonstrated for an infinitely long wire, to the finite-sized photonic nanowire SPS\cite{Gregersen:08,Gregersen:10, Claudon2010,Bleuse2011, Claudon13, Munsch2013,Bulgarini2014,Gregersen2016} sketched in Fig.\ (\ref{needle_eps}a). The central part consists of a finite-length nanowire segment, capped by a conical needle taper with half-opening angle $\theta$ and a total taper height of $h_{\rm taper}=D/(2\tan \theta)$. The purpose of the needle taper is to adiabatically expand the fundamental $\mathrm{HE}_{11}$ mode laterally to decrease the divergence angle and increase the collection efficiency\cite{Gregersen:08,Munsch2013,Bulgarini2014}. The tapering is also effective for coupling into a single-mode fiber\cite{Bulgarini2014}. We introduce rings with a flat top surface and height identical to the central nanowire segment. Increasing the ring height beyond this value could disrupt the adiabatic expansion, while a lower value closer to the QD position might diminish their beneficial effect. To reflect light propagating towards the substrate, we implement a bottom $\rm SiO_2$-$\rm Ag$ mirror\cite{Friedler:08} with a $\rm SiO_2$ layer thickness of $t_{\rm SiO_2}$. The QD is placed at a distance $h_{\rm b}$ ($h_{\rm t}$) from the bottom (top) interface, and we use $n_{\rm SiO_2}=1.4518$ and $n_{\rm Ag}=0.0747+6.4294i$.

 The physics of the photonic nanowire is in general well described \cite{Gregersen:10, Claudon2010, Claudon13, Munsch2013, Gregersen2016} using an SMM allowing for direct physical insight into the governing physics. In the following, we will compare simulations of the collection efficiency obtained using the full model, $\varepsilon_{g}(\mathrm{NA})=P_{g,\rm lens}(\mathrm{NA})/P_{\rm T}$, with the prediction, $\varepsilon_{g,\rm SMM}(\mathrm{NA})$, of the (nanowire) SMM given by
\begin{equation}
\varepsilon_{g,\rm SMM}(\mathrm{NA})=\gamma_{g}(\mathrm{NA})\frac{\beta_{\infty}(1+|r_{1,1,\mathrm{b}}|)^2}{2(1+\beta_{\infty}|r_{1,1,\mathrm{b}}|)},
\label{effSMM}
\end{equation} where $\gamma_{g}(\mathrm{NA})=P_{g,\rm HE_{11},lens}(\mathrm{NA})/P_{\rm HE_{11}}$ is the transmission coefficient of the fundamental $\mathrm{HE}_{11}$ mode taking into account an overlap with a Gaussian profile \cite{Munsch2013}, $\beta_{\infty}$ is the $\beta$ factor of the corresponding infinite-length nanowire structure, and $r_{1,1,\mathrm{b}}$ is the modal reflection of the fundamental $\mathrm{HE}_{11}$ mode at the bottom interface. In the derivation of the SMM, we assume that the background emission into radiation modes $P_{\rm R}$ of the finite structure is well approximated by that of the infinite structure, and we assume that the top reflectivity $r_{1,1,\rm t}$ at the interface between the nanowire segment and the taper is zero. 
We present the detailed dependence of $\gamma_{g}$ on the tapering angle $\theta$ and on the $\mathrm{NA}$ in Supplementary (S9), and we show that $r_{1,1,\rm t}$ indeed approaches 0 for small tapering angles. Furthermore, we show that the bottom reflectivity $r_{1,1,\rm b}$ is maximized at $t_{\rm SiO_2}=\SI{6}{nm}$ for all ring configurations.

To maximize the collection efficiency, the vertical position of the dipole should be aligned with the field antinode. Within the SMM, the antinode position is governed by the phase evolution of the fundamental $\mathrm{HE}_{11}$ mode. At the antinode, the top height $h_{\rm t}$ (bottom height $h_{\rm b}$) satisfies
\begin{equation}
n_{\rm j}2\pi=2\beta_1 h_{\rm j}+\arg(r_{1,1,\rm j}),\quad  n_{\rm j}\in \mathbb{N},
\label{zJ}
\end{equation} where $\beta_1$ is the propagation constant of the fundamental $\mathrm{HE}_{11}$ mode and j $\in$ (b,t). While $r_{1,1,\rm t} \sim 0$ for small tapering angles means there is no specific constraint on the top height, we have nonetheless used Eq.\ (\ref{zJ}) for all angles. The choice of integers $n_{\rm b}$ ($n_{\rm t}$) determines the number of antinodes between the bottom (top) interface and the dipole, and we have tested the combined set of $n_{\rm b,t}\in [1,2,3,4]$ resulting in a total of 16 combinations. The efficiency depends on the $n_{\rm b,t}$ combination as further discussed below. On the other hand, the SMM assumes a constant $P_{\rm R}$ given by the infinite structure and thus predicts no influence from $n_{\rm b,t}$ on $\varepsilon_{g,\rm SMM}(\mathrm{NA})$. 

\begin{figure}[h!]
	\begin{subfigure}{1\linewidth}
		\centering
		\includegraphics[width= 1 \linewidth]{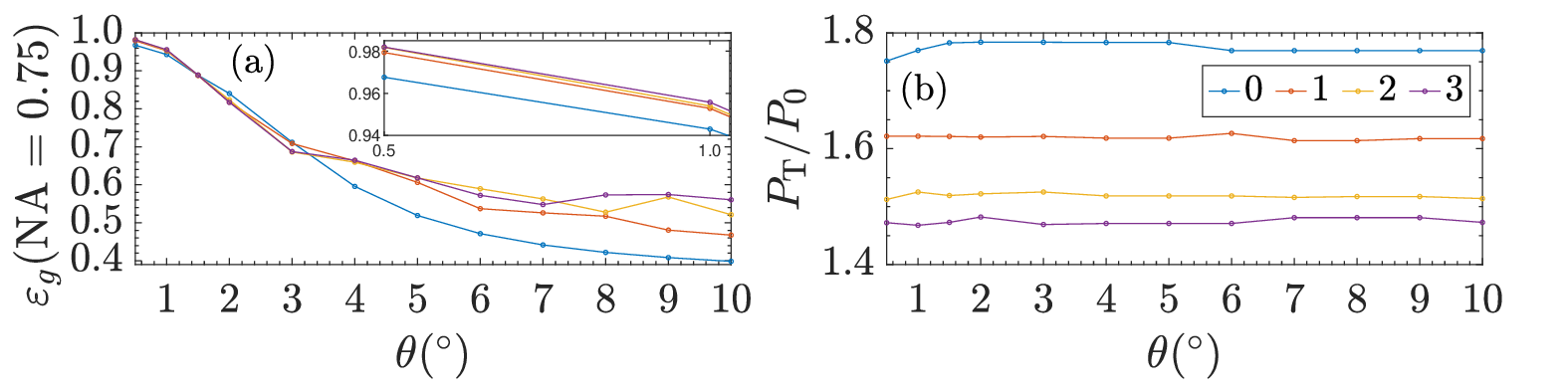}
	\end{subfigure}
	\caption{(a) Collection efficiency $\varepsilon_g (\mathrm{NA}=0.75)$ and (b) Purcell factor of the total emission $P_{\mathrm{T}}/P_{0}$ as a function of the tapering angle $\theta$ for 0-3 rings. $n_{\rm t}$ and $n_{\rm b}$ are chosen such that $\varepsilon_g (\mathrm{NA}=0.75)$ is maximized.}
	\label{Fptot_theta}
\end{figure}

We present calculated collection efficiency $\varepsilon_g (\mathrm{NA}=0.75)$ for the photonic nanowire SPS with 0-3 rings, using the optimized fully free parameters from Table \ref{table1}, in Fig.\ (\ref{needle_eps}b) as a function of the 16 combinations of $n_{\rm t}$ and $n_{\rm b}$ for a tapering angle of $\theta=0.5^{\circ}$ and an NA of 0.75. The dotted line is calculated using the SMM, while the solid line represents the full model, and we estimate an overall numerical uncertainty in the efficiency of 0.0010. Deviations between the full model and the SMM are observed for all configurations. The SMM generally slightly overestimates the performance, except for the $n_{\rm t}=3$ and $n_{\rm b}=2$ bare nanowire geometry, where a slight increase of 0.0075 is observed. Similar discrepancy between the SMM and the full model was also observed for the photonic "hourglass" SPS design \cite{Osterkryger2019} and the "nanopost" SPS geometry \cite{Jacobsen23}, which was attributed to the influence of the finite geometry on the emission $P_{\rm R}$ into background radiation modes, as well as coupling between the fundamental $\mathrm{HE}_{11}$ mode and the evanescent modes at the interfaces. Ultimately, the maximum collection efficiency as function of $n_{\rm b,t}$ is $\varepsilon_{g}=0.9676$ for the bare photonic nanowire SPS, $\varepsilon_{g}=0.9794$ for 1 ring, $\varepsilon_{g}=0.9818$ for 2 rings, and $\varepsilon_{g}=0.9820$ for 3 rings with a total increase of 0.0143 in the maximum collection efficiency from 0 to 3 rings. The maximum collection efficiencies are generally slightly lower than the maximum $\beta$ factors in the infinite structures, since both the bottom reflectivity and the top taper transmission coefficients take values below unity. Only in the bare wire case does $\varepsilon_{g}=0.9676$ (and also the SMM prediction $\varepsilon_{g,\rm SMM}=0.9596$) exceed $\beta_{\infty}=0.9581$ due to a slight benefit of the minor Purcell enhancement being present (see Fig.\ (\ref{Fptot_theta}b) and accompanying text). Nevertheless, for a multi-photon experiment with 50 photons, the success probability increases by a factor of $(0.9820/0.9676)^{50}\approx2$. This demonstrates that the efficiency increase obtained by suppressing the background emission using the rings can have a significant impact on the scale-up of multi-photon interference experiments.

The influence of the taper angle $\theta$ on the collection efficiency $\varepsilon_g (\mathrm{NA}=0.75)$ and the Purcell factor $P_{\mathrm{T}}/P_{0}$ is presented in Fig.\ (\ref{Fptot_theta}) for the four (0-3 rings) configurations considered. Here, $n_{\rm t}$ and $n_{\rm b}$ are chosen to maximize $\varepsilon_g (\mathrm{NA}=0.75)$. For all configurations, the efficiency increases with decreasing taper angle similarly to the transmission $\gamma_g (\mathrm{NA}=0.75)$ in Fig.\ (S9.2) in Supplementary (S9) thanks to the adiabatic expansion of the HE$_{11}$ mode \cite{Gregersen:08}.  For $\theta \leq 1 ^{\circ}$, where $\gamma_g\rightarrow1$ and $r_{1,1,\rm t}\rightarrow0$, the increase in the $\beta$ factor provided by the rings improves the efficiency. For larger taper angles, the transmission coefficient and the scattering effects are the dominating effects governing the collection efficiency. In this regime, significant reflection occurs from the top taper as shown in Fig.\ (S9.4) in Supplementary (S9) causing a significant breakdown of the SMM with increased influence of the vertical dipole position as shown in Fig.\ (S9.5) in Supplementary (S9). Therefore, comparably minor variations $\sim0.05$ of the $\beta$ factor in the infinite structures, for which the rings are solely optimized, are not the determining factor in the final collection efficiency for the larger taper angles.

The Purcell factor of the total emission $P_{\mathrm{T}}/P_{0}$ is shown in Fig.\ (\ref{Fptot_theta}b) as a function of the tapering angle $\theta$ for 0-3 rings, where $n_{\rm t}$ and $n_{\rm b}$ are chosen to maximize $\varepsilon_g (\mathrm{NA}=0.75)$. While $P_{\mathrm{T}}/P_{0}$ is hardly influenced by the tapering angle, it decreases as more rings are added. This occurs as the center diameter increases with the number of rings according to Table \ref{table1}, leading to a decrease in $P_{\mathrm{HE}_{11}}/P_0$ as shown in Fig.\ (\ref{beta_opt_2_3_combined}b). Nevertheless, the values of $P_{\mathrm{T}}/P_{0}$ in Fig.\ (\ref{Fptot_theta}b) are higher than for the corresponding infinite-length nanowire structures due to the minor Purcell enhancement provided by the highly reflective bottom mirror, which also serves to mitigate the effects of non-radiative recombination.

\section*{Micropillar} \label{Micropillar}
\begin{figure}[H]
	\begin{subfigure}{1\linewidth}
		\centering
		\includegraphics[width= 0.7 \linewidth]{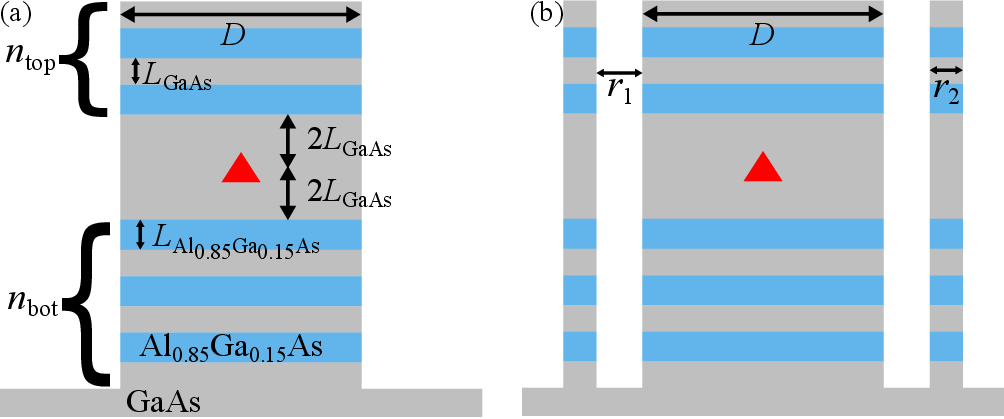}
	\end{subfigure}
	\caption{Sketch and geometrical parameters of (a) the bare micropillar SPS and (b) the geometry featuring 1 ring.}
	\label{micropillar_figure}
\end{figure} 

We now proceed to introduce the concept of suppression of radiation modes using rings to the micropillar SPS geometry illustrated in Fig.\ (\ref{micropillar_figure}). The micropillar SPS consists of a QD placed in the center of a vertical $\lambda$ cavity surrounded by distributed Bragg reflectors (DBRs) consisting of $n_{\rm top}$ ($n_{\rm bot}$) GaAs/$\rm Al_{0.85}Ga_{0.15}As$ layer pairs in the top (bottom) DBR. Whereas the photonic nanowire design approach relies predominantly on suppression of $P _{\rm R}$ to increase $\beta$ and the efficiency, the micropillar design relies primarily on Purcell enhancement enabled by the cavity effect to increase the two. Initial indications of suppression of the SE by placing rings around the micropillar have already been reported both numerically \cite{Blokhin:21} and experimentally \cite{Jakubczyk2014}.
Here, we build upon the state-of-the-art numerically optimized micropillar SPS \cite{Wang2020_PRB_Biying} and demonstrate that suppression of SE using rings can further improve the performance. 

\begin{figure}[h!]
	\begin{subfigure}{1\linewidth}
		\centering
		\includegraphics[width= 0.7 \linewidth]{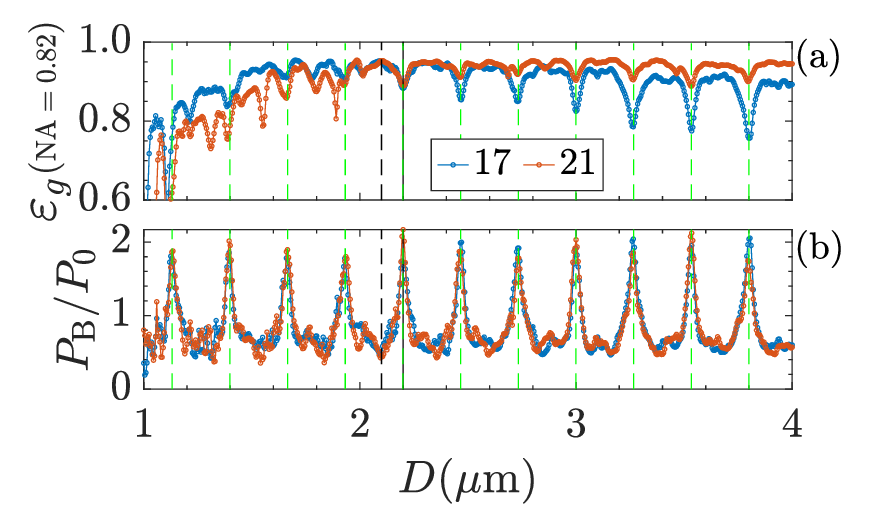} 
	\end{subfigure}
	\caption{(a) $\varepsilon_g (\mathrm{NA}=0.82)$ and (b) $P_{\mathrm{B}}/P_{0}$ as a function of the diameter, $D$, for $n_{\rm top}=17$ and $n_{\rm top}=21$ top DBR layer pairs respectively. The dotted green lines correspond to peaks in $P_{\mathrm{B}}$. The dotted black lines correspond to $D=\SI{2.1}{\mu m}$ and $D=\SI{2.2}{\mu m}$.} 
	\label{micropillar_standard_eff_PB}
\end{figure}

Our starting point is the bare micropillar SPS without rings optimized for collection efficiency and photon indistinguishability simultaneously\cite{Wang2020_PRB_Biying}, see Supplementary (S10) for geometrical parameters and details of the optimization procedure. In Fig.\ (\ref{micropillar_standard_eff_PB}), we present the power into background modes $P_{\rm B}/P_0$ and the efficiency $\varepsilon_g(\mathrm{NA}=0.82)$ of the bare optimized micropillar as a function of the diameter for $n_{\rm top}=17$ and 21 (see Supplementary Fig.\ (S10.2) for corresponding values of the $\beta$ factor, Purcell factor, etc.).
 The background emission $P_{\rm B}$ displays oscillatory variations with the diameter \cite{wang2021}, and we observe that peaks in $P_{\rm B}$ correspond to minima in $\varepsilon _g$. We now investigate the effect of adding initially a single ring for a diameter corresponding to a local maximum ($D=\SI{2.2}{\mu m}$) in $P_{\rm B}$ and subsequently a diameter corresponding to a local minimum ($D=\SI{2.1}{\mu m}$). Selecting the optimal ring parameters is not obvious\cite{Micro_ring_FDTD}, and unfortunately, we cannot use the infinite wire result, as $P_{\rm B}$ behaves differently in the micropillar, as further described in Supplementary (S10). In the following, we instead scan across the ($r_1$,$r_2$) parameter space. We use $\mathrm{NA}=0.82$ and again estimate an uncertainty of 0.0010 for all reported values of the collection efficiency.

\begin{figure}[h!]
	\begin{subfigure}{1\linewidth}
		\centering
		\includegraphics[width= 1 \linewidth]{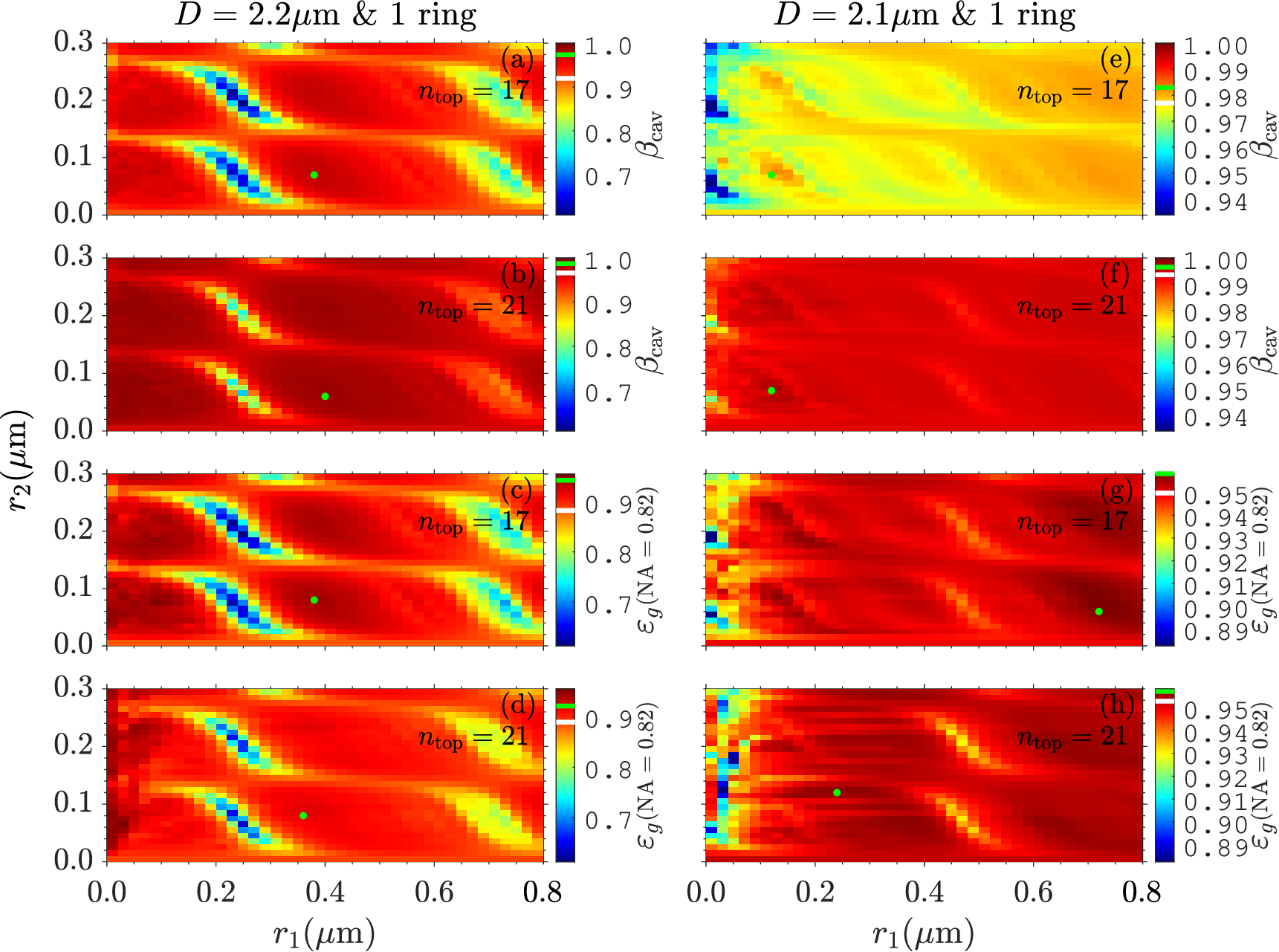}
	\end{subfigure}
	\caption{(a) $\beta_{\mathrm{cav}}$ with $n_{\rm top}=17$, (b) $\beta_{\mathrm{cav}}$ with $n_{\rm top}=21$, (c) $\varepsilon_g (\mathrm{NA}=0.82)$ with $n_{\rm top}=17$, and (d) $\varepsilon_g (\mathrm{NA}=0.82)$ with $n_{\rm top}=21$ as a function of the air gap, $r_1$, and the ring thickness, $r_2$, for 1 ring at $D=\SI{2.2}{\mu m}$. The green dots (a-d) indicate the maximum values with the restrictions of $r_1 > \SI{0.2}{\mu m}$ and $r_2<\SI{0.12}{\mu m}$. The white line in the colorbars (a-h) indicates the value without a ring and the green line indicates the value at the green dot. (e) $\beta_{\mathrm{cav}}$ with $n_{\rm top}=17$, (f) $\beta_{\mathrm{cav}}$ with $n_{\rm top}=21$, (g) $\varepsilon_g (\mathrm{NA}=0.82)$ with $n_{\rm top}=17$, and (h) $\varepsilon_g (\mathrm{NA}=0.82)$ with $n_{\rm top}=21$ as a function of the air gap, $r_1$, and the ring thickness, $r_2$, for 1 ring at $D=\SI{2.1}{\mu m}$. The green dots (e-h) indicate the maximum values without any restrictions. Note: a few of the darkest blue data points (e-h) have lower values than indicated by the color bars, but the limit is chosen to increase the visibility.}
	\label{micropillar_article_1ring_D2_2v3}
\end{figure} 

We present the calculated $\beta_{\mathrm{cav}}$ (we use the cav subscript when referring to the $\beta$ factor of the micropillar) in Figs.\ (\ref{micropillar_article_1ring_D2_2v3}a,b) and $\varepsilon_g$ in Figs.\ (\ref{micropillar_article_1ring_D2_2v3}c,d) for  $n_{\rm top}=17$ and 21 for a micropillar with 1 ring and a diameter $D=\SI{2.2}{\mu m}$. The first trend observed
is that overall, the variations of $\beta_{\mathrm{cav}}$ are also reflected in $\varepsilon_g$, indicating that $\beta_{\mathrm{cav}}$ plays a governing role in determining the efficiency, similar to the case of the bare micropillar. The green dots indicate the maximum values with the restrictions of $r_1 > \SI{0.2}{\mu m}$ and $r_2<\SI{0.12}{\mu m}$. For $n_{\rm top}=17$, the efficiency is increased from $\varepsilon_{g,17}=0.8829$ to $\varepsilon_{g,17,1}=0.9427$ (1 ring) with $r_1=\SI{0.38}{\mu m}$ and $r_2=\SI{0.08}{\mu m}$, and for $n_{\rm top}=21$ it is increased from $\varepsilon_{g,21}=0.8901
$ to $\varepsilon_{g,21,1}=0.9223$ with $r_1=\SI{0.36}{\mu m}$ and $r_2=\SI{0.08}{\mu m}$. The ring has thus increased the performance of the $D=\SI{2.2}{\mu m}$ micropillar to values approaching those of a bare micropillar with an optimal diameter. To identify the origin of this increase, we calculate and compare $\varepsilon_g$, $\varepsilon_{g,\mathrm{SMM}}$, $\beta_{\mathrm{cav}}$, and $\gamma_{g,\mathrm{cav}}$ in Table \ref{table2}. We observe that $\beta_{\mathrm{cav}}$ has indeed increased significantly, especially for $n_{\rm top}=17$, confirming that the ring has efficiently suppressed the background emission into the radiation modes and close-to-cutoff guided modes, which are at the origin of the limited efficiency of the $D=\SI{2.2}{\mu m}$ micropillar. This effect is also visible in the far-field pattern shown in Fig.\ (S10.3) in Supplementary (S10), which shows that the ring has suppressed the emission at high angles, originating from the peak in $P_{\rm B}$. However, we note that the improved efficiency is also partly resulting from an increase in $\gamma_{g,\mathrm{cav}}$, especially for $n_{\rm top}=21$. This increase may be caused by a beneficial scattering effect similar to the circular Bragg grating of the bullseye SPS\cite{Ates12}; however, further analysis of this effect is beyond the scope of this paper.

\begin{table*}[h!]
\centering
\begin{tabular}{|l|llll|llll|}
\hline
\multirow{3}{*}{}                                & \multicolumn{4}{c|}{$D=\SI{2.2}{\mu m}$}                                                                   & \multicolumn{4}{c|}{$D=\SI{2.1}{\mu m}$}                                                                   \\ \cline{2-9} 
                                                 & \multicolumn{2}{c|}{$n_{\mathrm{top}}=17$}                    & \multicolumn{2}{c|}{$n_{\mathrm{top}}=21$} & \multicolumn{2}{c|}{$n_{\mathrm{top}}=17$}                    & \multicolumn{2}{c|}{$n_{\mathrm{top}}=21$} \\ \cline{2-9} 
                                                 & \multicolumn{1}{l|}{Standard} & \multicolumn{1}{l|}{1 ring}   & \multicolumn{1}{l|}{Standard}  & 1 ring    & \multicolumn{1}{l|}{Standard} & \multicolumn{1}{l|}{1 ring}   & \multicolumn{1}{l|}{Standard}  & 1 ring    \\ \hline
$\varepsilon_{g}(\mathrm{NA}=0.82)$              & \multicolumn{1}{l|}{$0.8829$} & \multicolumn{1}{l|}{$0.9427$} & \multicolumn{1}{l|}{$0.8901$}  & $0.9223$  & \multicolumn{1}{l|}{$0.9497$} & \multicolumn{1}{l|}{$0.9582$} & \multicolumn{1}{l|}{$0.9529$}  & $0.9571$  \\ \hline
$\varepsilon_{g,\mathrm{SMM}}(\mathrm{NA}=0.82)$ & \multicolumn{1}{l|}{$0.8868$} & \multicolumn{1}{l|}{$0.9461$} & \multicolumn{1}{l|}{$0.8900$}  & $0.9243$  & \multicolumn{1}{l|}{$0.9501$} & \multicolumn{1}{l|}{$0.9572$} & \multicolumn{1}{l|}{$0.9496$}  & $0.9549$  \\ \hline
$\beta_{\mathrm{cav}}$                           & \multicolumn{1}{l|}{$0.9190$} & \multicolumn{1}{l|}{$0.9725$} & \multicolumn{1}{l|}{$0.9669$}  & $0.9876$  & \multicolumn{1}{l|}{$0.9769$} & \multicolumn{1}{l|}{$0.9816$} & \multicolumn{1}{l|}{$0.9937$}  & $0.9940$  \\ \hline
$\gamma_{g,\mathrm{cav}}(\mathrm{NA}=0.82)$                   & \multicolumn{1}{l|}{$0.9650$} & \multicolumn{1}{l|}{$0.9729$} & \multicolumn{1}{l|}{$0.9205$}  & $0.9359$  & \multicolumn{1}{l|}{$0.9726$} & \multicolumn{1}{l|}{$0.9751$} & \multicolumn{1}{l|}{$0.9556$}  & $0.9607$  \\ \hline
\end{tabular}
\caption{The parameters $\varepsilon_g(\mathrm{NA}=0.82)$, $\varepsilon_{g,\mathrm{SMM}}(\mathrm{NA}=0.82)$, $\beta_{\mathrm{cav}}$ and $\gamma_{g,\mathrm{cav}}(\mathrm{NA}=0.82)$ for the standard micropillar and the micropillar with 1 optimized ring with $n_{\rm top}=17$ or $n_{\rm top}=21$ at $D=\SI{2.2}{\mu m}$ and $D=\SI{2.1}{\mu m}$.}
\label{table2}
\end{table*}

We now present our results for the optimal $D=\SI{2.1}{\mu m}$ diameter corresponding to the local minimum in $P_{\rm B}$, and thus a local maximum in efficiency, in Figs.\ (\ref{micropillar_article_1ring_D2_2v3}e-\ref{micropillar_article_1ring_D2_2v3}h). Again, the variations in $\beta_{\mathrm{cav}}$ are reflected in the $\varepsilon_g$ plots,
however, with more apparent deviations, especially for $n_{\rm top}=21$. This indicates that variations in the efficiency are no longer predominantly due to variations in $\beta_{\mathrm{cav}}$ but also due to variations in $\gamma_{g,\mathrm{cav}}$ resulting from the presence of the ring. For this optimal diameter, the efficiency for $n_{\rm top}=17$ can be increased from $\varepsilon_{g,17}=0.9497$ to $\varepsilon_{g,17,1}=0.9582$ at $r_1=\SI{0.72}{\mu m}$ and $r_2=\SI{0.06}{\mu m}$ and for $n_{\rm top}=21$ from $\varepsilon_{g,21}=0.9529$ to $\varepsilon_{g,21,1}=0.9571$ at $r_1=\SI{0.24}{\mu m}$ and $r_2=\SI{0.12}{\mu m}$. Thus, the ring can still slightly improve the performance of the bare micropillar with an already optimal diameter. From Table \ref{table2}, $\beta_{\mathrm{cav}}$ appears to be the main origin for the increase in $\varepsilon_g$ for $n_{\rm top}=17$. However, for $n_{\rm top}=21$, the efficiency $\varepsilon_g$ is mainly limited by $\gamma_{g,\mathrm{cav}}$, not by $\beta_{\mathrm{cav}}$, and thus the observed increase in $\varepsilon_g$ is caused predominantly by a circular Bragg grating type of scattering effect, also observed at $D=\SI{2.2}{\mu m}$.

Additionally, we present the performance of a micropillar with $D=\SI{2.1}{\mu m}$ surrounded by two rings with fixed parameters in Fig.\ (S10.4) in Supplementary (S10). In this case, a maximum efficiency of $\varepsilon_{g,17,2}=0.9610$ at $r_1=\SI{0.13}{\mu m}$ and $r_2=\SI{0.08}{\mu m}$ is obtained, representing a 0.008 increase from $\varepsilon_{g,21}=0.9529$ for the maximum bare pillar efficiency to $\varepsilon_{g,17,2}=0.9610$ with two rings.

Finally, we note that the micropillar SPS is less vulnerable to non-radiative recombination effects because of the high Purcell factors. Furthermore, for a dipole displacement up to $\SI{200}{nm}$, the Purcell factor is approximately reduced by $8\%$ and the collection only by $2\%$\cite{Madigawa2024}, showcasing the robustness of the micropillar.

\section*{Discussion}

Designing an SPS with a collection efficiency approaching unity requires suppression of all loss channels. The single-mode model $\varepsilon _{g,\mathrm{SMM}} = \gamma _g\beta$ allows a description of the efficiency as a product of the SE $\beta$ factor from the QD to the cavity mode and the subsequent transmission $\gamma _g$ to the collection optics. For the infinite wire geometry, we demonstrate that the photonic bandgap provided by a few (1-3) concentric rings can strongly suppress spontaneous emission into the radiation modes, allowing for a $\beta$ factor above 0.999 for an optimum diameter.

We initially applied the photonic bandgap concept to the photonic nanowire SPS operating in the single-guided-mode regime, for which radiation modes constitute the main loss mechanism. The physics of the photonic nanowire is generally well described by the SMM in Eq.\ \eqref{effSMM}, and the result of the infinite wire could directly be applied in the small-taper-angle regime. While $\beta$ exceeds 0.999 in the presence of rings, a lower collection efficiency of $\varepsilon _g \sim 0.98$ was obtained. The reduced total efficiency is a result of other loss channels of the finite photonic nanowire geometry, which become dominating in the near-unity regime. Thus, further increase of the efficiency will require optimization of the taper transmission $\gamma_g$ and the reflectivity $r_{1,1,\rm b}$ of the bottom mirror. While the SMM in Eq.\ \eqref{effSMM} generally provides a good description of the efficiency, it relies on the assumptions that $P_{\rm B}$ into radiation modes in the finite photonic nanowire is identical to $P_{\rm B}$ of the infinite wire and that only the HE$_{11}$ mode emission channel contributes to the collection in the far field. When placing the dipole at different vertical field antinodes, we observe deviations between the exact simulation and the SMM, particularly for larger angles. These deviations indicate that the emission into radiation modes and its suppression by the rings could be influenced by the presence of the bottom mirror, which could also direct light towards the collection optics through additional channels \cite{Jacobsen23}. This leaves room for further optimization of the height of the straight nanowire section and the position of the dipole beyond the SMM. Additionally, we note that the optimal center nanowire diameter, allowing for a maximum $\beta$ factor, increases slightly for each added ring. While this diameter increase leads to slightly longer taper for the needle tapering strategy \cite{Gregersen:08,Claudon2010} considered here, it, on the other hand, leads to a reduction of the taper height for the inverse "trumpet" taper strategy \cite{Gregersen:10,Munsch2013,Stepanov15,Gregersen2016,Cadeddu16}. 

We subsequently implemented the photonic bandgap concept for the micropillar SPS operating in the $\sim 2$ $\mu$m diameter multi-mode regime, where higher-order guided modes are present. Here, a major part of the background emission in the micropillar is into higher-order guided modes, and the rings offer no control of the SE into these modes. As a consequence, we cannot expect $P_{\rm B}$ to be suppressed to the same level as in the small diameter range of the infinite wire. Nevertheless, the introduction of rings still significantly impacts the background emission. At the suboptimal diameter $D=2.2 \mu$m, corresponding to a peak in $P_{\rm B}$, a single ring can suppress $P_{\rm B}$ and increase the efficiency $\varepsilon _g$ from 0.8829 to 0.9427 almost reaching the performance of 0.9497 for the optimal diameter $D=2.1 \mu$m of the bare $n_{\rm top} = 17$ micropillar. Further performance increase is limited by our lack of control of the emission $P_{\rm HG} \neq 0$ into higher-order guided modes. 
Nevertheless, we observed that the introduction of 2 rings to the optimum $D=2.1 \mu$m micropillar SPS allowed for an increase in efficiency from $\varepsilon_{g,21} = 0.9529$ for the bare micropillar to $\varepsilon_{g,17,2} = 0.9610$ with two rings. While this demonstrates that adding rings can indeed improve the performance of the bare micropillar SPS, we note that the introduction of the rings increases not only $\beta_{\rm cav}$ but also $\gamma _g$, indicating that the rings serve not only to suppress radiation modes but also to scatter light towards the collection optics similar to the circular Bragg grating in the Bullseye design. Further analysis of this effect and additional optimization of the micropillar is beyond the scope of the present paper and is left for future work.

While the nanowire and the micropillar with rings may appear conceptually similar to the circular Bragg grating in the bullseye design, the purpose of the rings is different. In the bullseye design, the purpose is to create a resonance effect to increase the Purcell factor and to scatter light upward to increase collection efficiency\cite{Ates12}, which generally requires a larger number of rings (>3). Therefore, the physics cannot be directly compared between the bullseye design and the cases considered in this paper, even if the rings may have a positive scattering effect in the micropillar case.

In this paper, we have only considered the designs for the purpose of single photon generation; however, both the photonic nanowire\cite{Versteegh2014} and the micropillar\cite{gines2022} have been used for entangled photon pair generation, and the increased collection efficiency will also be beneficial in this case. As mentioned in the introduction, the designs incorporating background suppression typically provide broadband efficiency, and we expect the suppression effect provided by the rings to be similar, which is particularly beneficial for entangled photon pair generation via the biexciton-exciton cascade\cite{gines2022}.

While we have fixed the wavelength and materials in this paper, we will also note that the suppression effect of the rings can be achieved across other material platforms with different refractive index. However, the optimization procedure will be case specific.

Even with a theoretical SPS design of near-unity efficiency featuring suppression of all loss channels, fabrication imperfections can be a bottleneck and such fabrication challenges must be overcome to realize the perfect SPS. We identify two other remaining bottlenecks in the quest for highly efficient sources of indistinguishable single photons, namely, (1) the state preparation scheme and (2) coupling to a single-mode fiber.
First, in the current work, we have modeled the emitter as a classical electric point dipole. However, from a quantum-mechanical perspective, the emitter is a two-level system that must initially be in the excited state. This calls for a state preparation protocol that initializes the excited state on demand with near-unity fidelity and on a timescale much shorter than the lifetime (i.e. of a few picoseconds). Above-band excitation is a common method that allows straightforward filtering of the pumping laser. However, it generates significant time-jitter and charge noise, which in turn lowers the indistinguishability\cite{Huber_2015}. Resonant excitation, on the other hand, yields excellent indistinguishability (>99$\%$), but requires a cross-polarization setup that reduces the total system efficiency by at least a factor of 2. While novel off-resonant excitation schemes have been proposed in recent years\cite{Thomas2021,Karli2022,Piccinini2025,Sbresny2022,Wei2022,Yan2022}, identifying a scheme that allows for simultaneous near-unity efficiency and indistinguishability remains an open research problem.
Second, while we have considered the collection efficiency to a first lens in this paper, the emitted photons should ideally be coupled to a single-mode fiber, which remains a challenge. Here, the most promising solution appears to glue the fiber to the top facet of the source, or very close to it\cite{Rickert2021,Margaria2025}. Recently, a fiber-pigtailed device with an efficiency of 20.8$\%$ at the output fiber has been demonstrated\cite{Margaria2025}. However, achieving 100$\%$ transmission between the cavity mode and a single-mode fiber is a challenging problem that requires additional investigation.

\section*{Conclusion}
We have provided a detailed physical description of how the introduction of concentric rings around an infinite wire influences the emission rates into both guided and radiation modes. Our infinite wire simulations were performed using a fully analytical method allowing for direct analysis of the physics of emission into radiation modes as well as fast computation speed allowing exploration of a large parameter space. We then demonstrated that 1-3 rings could further suppress the emission into radiation modes due to a photonic bandgap effect, achieving $\beta$ factors up to 0.999. This result was then applied to the photonic nanowire SPS, which allowed an increase in the maximum collection efficiency from $\varepsilon_{g,0}=0.9676$ without any rings to $\varepsilon_{g,3}=0.9820$ with three rings, further pushing the theoretical limit towards near-unity collection efficiency. This increase in collection efficiency corresponds to an increase in the success probability of a multi-photon interference experiment with 50 photons by a factor of $(0.9820/0.9676)^{50}\approx2$. We also introduced rings around the micropillar SPS, resulting in an increase in the collection efficiency from $\varepsilon_{g,17}=0.9529$ for the bare optimized micropillar to $\varepsilon_{g,17,2}=0.9610$ with two rings. We have thus demonstrated how rings improve upon two well established SPS designs.

\section*{Data availability}
The datasets used and/or analyzed during the current study are available upon reasonable request from the corresponding author.

\bibliography{sample}

@article{Somaschi2016,
	author = {Somaschi, N. and Giesz, V. and {De Santis}, L. and Loredo, J. C. and Almeida, M. P. and Hornecker, G. and Portalupi, S. L. and Grange, T. and Ant{\'{o}}n, C. and Demory, J. and G{\'{o}}mez, C. and Sagnes, I. and Lanzillotti-Kimura, N. D. and Lema{\'{i}}tre, A. and Auffeves, A. and White, A. G. and Lanco, L. and Senellart, P.},
	doi = {10.1038/nphoton.2016.23},
	issn = {1749-4885},
	journal = {Nat. Photonics},
	month = {mar},
	number = {5},
	pages = {340--345},
	publisher = {Nature Publishing Group},
	shorttitle = {Nat Photon},
	title = {{Near-optimal single-photon sources in the solid state}},
	url = {http://dx.doi.org/10.1038/nphoton.2016.23},
	volume = {10},
	year = {2016}
}

@ARTICLE{Tomm2021,
	author = {{Tomm}, Natasha and {Javadi}, Alisa and {Antoniadis}, Nadia O. and {Najer}, Daniel and {L{\"o}bl}, Matthias C. and {Korsch}, Alexander R. and {Schott}, R{\"u}diger and {Valentin}, Sascha R. and {Wieck}, Andreas D. and {Ludwig}, Arne and {Warburton}, Richard J.},
	title = "{A bright and fast source of coherent single photons}",
	journal = {Nat. Nanotechnol.},
	volume = {16},
	pages = {399–403},
	year = {2021},
	doi = {10.1038/s41565-020-00831-x},
}

@article{Wang2020_PRB_Biying,
	title = {Micropillar single-photon source design for simultaneous near-unity efficiency and indistinguishability},
	author = {Wang, Bi-Ying and Denning, Emil V. and G\"ur, U\u{g}ur Meri{\c{c}} and Lu, Chao-Yang and Gregersen, Niels},
	journal = {Phys. Rev. B},
	volume = {102},
	issue = {12},
	pages = {125301},
	numpages = {10},
	year = {2020},
	month = {Sep},
	publisher = {American Physical Society},
	doi = {10.1103/PhysRevB.102.125301},
	url = {https://link.aps.org/doi/10.1103/PhysRevB.102.125301}
}

@book{novotny2012principles,
author = {Novotny, Lukas and Hecht, Bert},
publisher = {Cambridge University Press},
title = {{Principles of Nano-Optics}},
year = {2012}
}

@book{NUMERICALMETHOD2014,
title = "Numerical Methods in Photonics",
author = "Andrei Lavrinenko and Jesper L{\ae}gsgaard and Niels Gregersen and Frank Schmidt and Thomas S{\o}ndergaard",
year = "2014",
isbn = "9781466563889",
publisher = "CRC Press",
}

@book{Yariv_6th,
    author = {{Yariv}, Amnon and {Yeh}, Pochi},
    title = "{Photonics: Optical Electronics in Modern Communications }",
    publisher = {Oxford University Press},
    edition = {6th},
    year = {2007},
}

@article{Friedler:08,
author = {I. Friedler and P. Lalanne and J. P. Hugonin and J. Claudon and J. M. G\'{e}rard and A. Beveratos and I. Robert-Philip},
journal = {Opt. Lett.},
number = {22},
pages = {2635--2637},
publisher = {Optica Publishing Group},
title = {Efficient photonic mirrors for semiconductor nanowires},
volume = {33},
month = {Nov},
year = {2008},
url = {http://opg.optica.org/ol/abstract.cfm?URI=ol-33-22-2635},
doi = {10.1364/OL.33.002635},
}

@Article{Jacobsen23,
author ="Jacobsen, Martin Arentoft and Wang, Yujing and Vannucci, Luca and Claudon, Julien and Gérard, Jean-Michel and Gregersen, Niels",
title  ="Performance of the nanopost single-photon source: beyond the single-mode model",
journal  ="Nanoscale",
year  ="2023",
volume  ="15",
issue  ="13",
pages  ="6156-6169",
publisher  ="The Royal Society of Chemistry",
doi  ="10.1039/D2NR07132K",
url  ="http://dx.doi.org/10.1039/D2NR07132K"}

@article{Zhou_2022,
doi = {10.1088/2058-9565/ac5918},
url = {https://dx.doi.org/10.1088/2058-9565/ac5918},
year = {2022},
month = {mar},
publisher = {IOP Publishing},
volume = {7},
number = {2},
pages = {025023},
author = {Xiaoyan Zhou and Peter Lodahl and Leonardo Midolo},
title = {In-plane resonant excitation of quantum dots in a dual-mode photonic-crystal waveguide with high $\beta$-factor},
journal = {Quantum Science and Technology}
}

@article{Beatrice22,
author = {Da Lio, Beatrice and Faurby, Carlos and Zhou, Xiaoyan and Chan, Ming Lai and Uppu, Ravitej and Thyrrestrup, Henri and Scholz, Sven and Wieck, Andreas D. and Ludwig, Arne and Lodahl, Peter and Midolo, Leonardo},
title = {A Pure and Indistinguishable Single-Photon Source at Telecommunication Wavelength},
journal = {Advanced Quantum Technologies},
volume = {5},
number = {5},
pages = {2200006},
doi = {https://doi.org/10.1002/qute.202200006},
url = {https://onlinelibrary.wiley.com/doi/abs/10.1002/qute.202200006},
year = {2022}
}

@article{Pan2012,
  title = {Multiphoton entanglement and interferometry},
  author = {Pan, Jian-Wei and Chen, Zeng-Bing and Lu, Chao-Yang and Weinfurter, Harald and Zeilinger, Anton and \ifmmode \dot{Z}\else \.{Z}\fi{}ukowski, Marek},
  journal = {Rev. Mod. Phys.},
  volume = {84},
  issue = {2},
  pages = {777--838},
  numpages = {0},
  year = {2012},
  month = {May},
  publisher = {American Physical Society},
  doi = {10.1103/RevModPhys.84.777},
  url = {https://link.aps.org/doi/10.1103/RevModPhys.84.777}
}

@article{OBrien2009,
author = {O'Brien, Jeremy L. and Furusawa, Akira and Vu{\v{c}}kovi{\'{c}}, Jelena},
doi = {10.1038/nphoton.2009.229},
isbn = {1749-4885},
issn = {1749-4885},
journal = {Nat. Photonics},
number = {12},
pages = {687--695},
publisher = {Nature Publishing Group},
title = {{Photonic quantum technologies}},
volume = {3},
year = {2009}
}

@article{Huber2018,
author = {Huber, Daniel and Reindl, Marcus and Aberl, Johannes and Rastelli, Armando and Trotta, Rinaldo},
doi = {10.1088/2040-8986/aac4c4},
journal = {J. Opt.},
month = {jul},
number = {7},
pages = {073002},
publisher = {IOP Publishing},
title = {{Semiconductor quantum dots as an ideal source of polarization-entangled photon pairs on-demand: a review}},
url = {http://stacks.iop.org/2040-8986/20/i=7/a=073002?key=crossref.d887ba042c347e39494c3d66fbbe9372},
volume = {20},
year = {2018}
}

@article{Aharonovich2016,
author = {Aharonovich, Igor and Englund, Dirk and Toth, Milos},
doi = {10.1038/nphoton.2016.186},
issn = {1749-4885},
journal = {Nat. Photonics},
number = {10},
pages = {631--641},
publisher = {Nature Publishing Group},
title = {{Solid-state single-photon emitters}},
url = {http://www.nature.com/doifinder/10.1038/nphoton.2016.186},
volume = {10},
year = {2016}
}

@article{Gregersen2013,
author = {Gregersen, Niels and Kaer, Per and M{\o}rk, Jesper},
doi = {10.1109/JSTQE.2013.2255265},
issn = {1077-260X},
journal = {IEEE J. Sel. Top. Quantum Electron.},
number = {5},
pages = {9000516},
title = {{Modeling and Design of High-Efficiency Single-Photon Sources}},
url = {http://ieeexplore.ieee.org/lpdocs/epic03/wrapper.htm?arnumber=6493378},
volume = {19},
year = {2013}
}

@incollection{Gregersen2017,
address = {Boca Raton},
author = {Gregersen, Niels and McCutcheon, Dara P. S. and M{\o}rk, Jesper},
booktitle = {Handbook of Optoelectronic Device Modeling and Simulation Vol. 2},
chapter = {46},
editor = {Piprek, Joachim},
pages = {585--607},
publisher = {CRC Press},
title = {{Single-Photon Sources}},
year = {2017}
}

@article{Shields2007,
author={Shields, Andrew J.},
title={Semiconductor quantum light sources},
journal={Nature Photonics},
year={2007},
month={Apr},
day={01},
volume={1},
number={4},
pages={215-223},
issn={1749-4893},
doi={10.1038/nphoton.2007.46},
url={https://doi.org/10.1038/nphoton.2007.46}
}

@article{Purcell1946,
author = {Purcell, E. M.},
journal = {Phys. Rev.},
pages = {681},
title = {{Spontaneous emission probabilities at radio frequencies}},
volume = {69},
year = {1946}
}

@article{Wang2019b,
author = {Wang, Hui and He, Yu-Ming and Chung, T.-H. and Hu, Hai and Yu, Ying and Chen, Si and Ding, Xing and Chen, M.-C. and Qin, Jian and Yang, Xiaoxia and Liu, Run-Ze and Duan, Z.-C. and Li, J.-P. and Gerhardt, S. and Winkler, K. and Jurkat, J. and Wang, Lin-Jun and Gregersen, Niels and Huo, Yong-Heng and Dai, Qing and Yu, Siyuan and H{\"{o}}fling, Sven and Lu, Chao-Yang and Pan, Jian-Wei},
doi = {10.1038/s41566-019-0494-3},
journal = {Nat. Photonics},
month = {nov},
number = {11},
pages = {770--775},
publisher = {Nature Publishing Group},
title = {{Towards optimal single-photon sources from polarized microcavities}},
url = {http://www.nature.com/articles/s41566-019-0494-3},
volume = {13},
year = {2019}
}

@article{Wang2019a,
  title = {{On-Demand Semiconductor Source of Entangled Photons Which Simultaneously Has High Fidelity, Efficiency, and Indistinguishability}},
  author = {Wang, Hui and Hu, Hai and Chung, T.-H. and Qin, Jian and Yang, Xiaoxia and Li, J.-P. and Liu, R.-Z. and Zhong, H.-S. and He, Y.-M. and Ding, Xing and Deng, Y.-H. and Dai, Qing and Huo, Y.-H. and H\"ofling, Sven and Lu, Chao-Yang and Pan, Jian-Wei},
  journal = {Phys. Rev. Lett.},
  volume = {122},
  issue = {11},
  pages = {113602},
  numpages = {6},
  year = {2019},
  month = {Mar},
  publisher = {American Physical Society},
  doi = {10.1103/PhysRevLett.122.113602},
  url = {https://link.aps.org/doi/10.1103/PhysRevLett.122.113602}
}

@article{Rickert:19,
author = {Lucas Rickert and Timm Kupko and Sven Rodt and Stephan Reitzenstein and Tobias Heindel},
journal = {Opt. Express},
number = {25},
pages = {36824--36837},
publisher = {Optica Publishing Group},
title = {Optimized designs for telecom-wavelength quantum light sources based on hybrid circular {B}ragg gratings},
volume = {27},
month = {Dec},
year = {2019},
url = {https://opg.optica.org/oe/abstract.cfm?URI=oe-27-25-36824},
doi = {10.1364/OE.27.036824},
}

@ARTICLE{Ates12,
  author={Ates, Serkan and Sapienza, Luca and Davanco, Marcelo and Badolato, Antonio and Srinivasan, Kartik},
  journal={IEEE Journal of Selected Topics in Quantum Electronics}, 
  title={{Bright Single-Photon Emission From a Quantum Dot in a Circular Bragg Grating Microcavity}}, 
  year={2012},
  volume={18},
  number={6},
  pages={1711-1721},
  doi={10.1109/JSTQE.2012.2193877}}

@article{Davanço11,
    author = {Davanço, M. and Rakher, M. T. and Schuh, D. and Badolato, A. and Srinivasan, K.},
    title = "{A circular dielectric grating for vertical extraction of single quantum dot emission}",
    journal = {Applied Physics Letters},
    volume = {99},
    number = {4},
    pages = {041102},
    year = {2011},
    month = {07},
    issn = {0003-6951},
    doi = {10.1063/1.3615051},
    url = {https://doi.org/10.1063/1.3615051},
}

@article{Dusanowska20,
author = {Mocza\l{}a-Dusanowska, Magdalena and Dusanowski, \L{}ukasz and Iff, Oliver and Huber, Tobias and Kuhn, Silke and Czyszanowski, Tomasz and Schneider, Christian and Höfling, Sven},
title = {{Strain-Tunable Single-Photon Source Based on a Circular Bragg Grating Cavity with Embedded Quantum Dots}},
journal = {ACS Photonics},
volume = {7},
number = {12},
pages = {3474-3480},
year = {2020},
doi = {10.1021/acsphotonics.0c01465},

URL = { 
        https://doi.org/10.1021/acsphotonics.0c01465
    
},
}

@article{Ding2016,
author = {Ding, Xing and He, Yu and Duan, Z.-C. and Gregersen, Niels and Chen, M.-C. and Unsleber, S. and Maier, S. and Schneider, Christian and Kamp, Martin and H{\"{o}}fling, Sven and Lu, Chao-Yang and Pan, Jian-Wei},
doi = {10.1103/PhysRevLett.116.020401},
issn = {10797114},
journal = {Phys. Rev. Lett.},
number = {2},
pages = {020401},
title = {{On-Demand Single Photons with High Extraction Efficiency and Near-Unity Indistinguishability from a Resonantly Driven Quantum Dot in a Micropillar}},
url = {http://arxiv.org/abs/1601.00284},
volume = {116},
year = {2016}
}

@article{Bleuse2011,
author = {Bleuse, Jo{\"{e}}l and Claudon, Julien and Creasey, Megan and Malik, Nitin S. and G{\'{e}}rard, Jean-Michel and Maksymov, Ivan and Hugonin, Jean-Paul and Lalanne, Philippe},
doi = {10.1103/PhysRevLett.106.103601},
isbn = {00319007},
issn = {00319007},
journal = {Phys. Rev. Lett.},
number = {10},
pages = {103601},
pmid = {21469790},
title = {{Inhibition, enhancement, and control of spontaneous emission in photonic nanowires}},
volume = {106},
year = {2011}
}

@article{Claudon2010,
author = {Claudon, Julien and Bleuse, Jo{\"{e}}l and Malik, Nitin Singh and Bazin, Maela and Jaffrennou, P{\'{e}}rine and Gregersen, Niels and Sauvan, Christophe and Lalanne, Philippe and G{\'{e}}rard, Jean-Michel},
doi = {10.1038/nphoton.2009.287},
issn = {1749-4885},
journal = {Nat. Photonics},
month = {feb},
number = {3},
pages = {174--177},
publisher = {Nature Publishing Group},
shorttitle = {Nat Photon},
title = {{A highly efficient single-photon source based on a quantum dot in a photonic nanowire}},
url = {http://dx.doi.org/10.1038/nphoton.2009.287},
volume = {4},
year = {2010}
}

@article{Gregersen2016,
author = {Gregersen, Niels and McCutcheon, Dara P. S. and M{\o}rk, Jesper and G{\'{e}}rard, Jean-Michel and Claudon, Julien},
doi = {10.1364/OE.24.020904},
journal = {Opt. Express},
month = {sep},
number = {18},
pages = {20904--20924},
publisher = {Optical Society of America},
title = {{A broadband tapered nanocavity for efficient nonclassical light emission}},
url = {https://www.osapublishing.org/abstract.cfm?URI=oe-24-18-20904},
volume = {24},
year = {2016}
}

@article{Gaal2022,
author = {Ga{\'{a}}l, Benedek and Vannucci, Luca and Jacobsen, Martin Arentoft and Claudon, Julien and G{\'{e}}rard, Jean-Michel and Gregersen, Niels},
title = {Near-unity efficiency and photon indistinguishability for the “hourglass” single-photon source using suppression of the background emission},
journal = {Applied Physics Letters},
volume = {121},
number = {17},
pages = {170501},
year = {2022},
doi = {10.1063/5.0107624}
}

@article{Arcari2014,
  title = {Near-{U}nity {C}oupling {E}fficiency of a {Q}uantum {E}mitter to a {P}hotonic {C}rystal {W}aveguide},
  author = {Arcari, M. and S\"ollner, I. and Javadi, A. and Lindskov Hansen, S. and Mahmoodian, S. and Liu, J. and Thyrrestrup, H. and Lee, E. H. and Song, J. D. and Stobbe, S. and Lodahl, P.},
  journal = {Phys. Rev. Lett.},
  volume = {113},
  issue = {9},
  pages = {093603},
  numpages = {5},
  year = {2014},
  month = {Aug},
  publisher = {American Physical Society},
  doi = {10.1103/PhysRevLett.113.093603},
  url = {https://link.aps.org/doi/10.1103/PhysRevLett.113.093603}
}

@article{Lecamp2007b,
author = {Lecamp, G. and Lalanne, P. and Hugonin, J. P.},
doi = {10.1103/PhysRevLett.99.023902},
isbn = {0031-9007 (Print)$\backslash$r0031-9007 (Linking)},
issn = {00319007},
journal = {Phys. Rev. Lett.},
number = {2},
pages = {023902},
pmid = {17678224},
title = {{Very {L}arge {S}pontaneous-{E}mission $\beta$ {F}actors in {P}hotonic-{C}rystal {W}aveguides}},
volume = {99},
year = {2007}
}

@article{MangaRao2007,
author = {{Manga Rao}, V. S. C. and Hughes, S.},
doi = {10.1103/PhysRevB.75.205437},
isbn = {1098-0121$\backslash$n1550-235X},
issn = {10980121},
journal = {Phys. Rev. B},
number = {20},
pages = {205437},
title = {{Single quantum-dot {P}urcell factor and $\beta$ factor in a photonic crystal waveguide}},
volume = {75},
year = {2007}
}

@article{Yao2018,
author = {Yao, Beimeng and Su, Rongbin and Wei, Yuming and Liu, Zhuojun and Zhao, Tianming and Liu, Jin},
doi = {10.3938/jkps.73.1502},
journal = {J. Korean Phys. Soc.},
month = {nov},
number = {10},
pages = {1502--1505},
publisher = {The Korean Physical Society},
title = {{Design for {H}ybrid {C}ircular {B}ragg {G}ratings for a {H}ighly {E}fficient {Q}uantum-{D}ot {S}ingle-{P}hoton {S}ource}},
url = {http://link.springer.com/10.3938/jkps.73.1502},
volume = {73},
year = {2018}
}

@article{Liu2019,
author = {Liu, Jin and Su, Rongbin and Wei, Yuming and Yao, Beimeng and da Silva, Saimon Filipe Covre and Yu, Ying and Iles-Smith, Jake and Srinivasan, Kartik and Rastelli, Armando and Li, Juntao and Wang, Xuehua},
doi = {10.1038/s41565-019-0435-9},
journal = {Nat. Nanotechnol.},
month = {jun},
number = {6},
pages = {586--593},
publisher = {Nature Publishing Group},
title = {{A solid-state source of strongly entangled photon pairs with high brightness and indistinguishability}},
url = {http://www.nature.com/articles/s41565-019-0435-9},
volume = {14},
year = {2019}
}

@article{Kotal2021,
author = {Kotal, Saptarshi and Artioli, Alberto and Wang, Yujing and Osterkryger, Andreas Dyhl and Finazzer, Matteo and Fons, Romain and Genuist, Yann and Bleuse, Jo{\"{e}}l and G{\'{e}}rard, Jean-Michel and Gregersen, Niels and Claudon, Julien},
doi = {10.1063/5.0045834},
issn = {0003-6951},
journal = {Appl. Phys. Lett.},
month = {may},
number = {19},
pages = {194002},
publisher = {AIP Publishing LLC AIP Publishing},
title = {{A nanowire optical nanocavity for broadband enhancement of spontaneous emission}},
url = {https://aip.scitation.org/doi/10.1063/5.0045834},
volume = {118},
year = {2021}
}

@article{Friedler2009,
author = {I. Friedler and C. Sauvan and J. P. Hugonin and P. Lalanne and J. Claudon and J. M. G\'{e}rard},
journal = {Opt. Express},
number = {4},
pages = {2095--2110},
publisher = {Optica Publishing Group},
title = {Solid-state single photon sources: the nanowire antenna},
volume = {17},
month = {Feb},
year = {2009},
url = {https://opg.optica.org/oe/abstract.cfm?URI=oe-17-4-2095},
doi = {10.1364/OE.17.002095},
}

@article{Osterkryger2019,
author = {Osterkryger, Andreas Dyhl and G{\'{e}}rard, Jean-Michel and Claudon, Julien and Gregersen, Niels},
doi = {10.1364/OL.44.002617},
journal = {Opt. Lett.},
month = {jun},
number = {11},
pages = {2617--2620},
publisher = {Optical Society of America},
title = {{Photonic “hourglass” design for efficient quantum light emission}},
url = {https://www-osapublishing-org.proxy.findit.dtu.dk/ol/abstract.cfm?uri=ol-44-11-2617},
volume = {44},
year = {2019}
}

@article{Arakawa2020,
author = {Arakawa, Yasuhiko and Holmes, Mark J.},
doi = {10.1063/5.0010193},
journal = {Appl. Phys. Rev.},
month = {jun},
number = {2},
pages = {021309},
publisher = {AIP Publishing LLC AIP Publishing},
title = {{Progress in quantum-dot single photon sources for quantum information technologies: {A} broad spectrum overview}},
url = {http://aip.scitation.org/doi/10.1063/5.0010193},
volume = {7},
year = {2020}
}

@article{Munsch2013,
  title = {Dielectric {G}a{A}s {A}ntenna {E}nsuring an {E}fficient {B}roadband {C}oupling between an {I}n{A}s {Q}uantum {D}ot and a {G}aussian {O}ptical {B}eam},
  author = {Munsch, Mathieu and Malik, Nitin S. and Dupuy, Emmanuel and Delga, Adrien and Bleuse, Jo\"el and G\'erard, Jean-Michel and Claudon, Julien and Gregersen, Niels and M\o{}rk, Jesper},
  journal = {Phys. Rev. Lett.},
  volume = {110},
  issue = {17},
  pages = {177402},
  numpages = {5},
  year = {2013},
  month = {Apr},
  publisher = {American Physical Society},
  doi = {10.1103/PhysRevLett.110.177402},
  url = {https://link.aps.org/doi/10.1103/PhysRevLett.110.177402}
}

@article{JMG1998,
  title = {Enhanced {S}pontaneous {E}mission by {Q}uantum {B}oxes in a {M}onolithic {O}ptical {M}icrocavity},
  author = {G\'erard, J. M. and Sermage, B. and Gayral, B. and Legrand, B. and Costard, E. and Thierry-Mieg, V.},
  journal = {Phys. Rev. Lett.},
  volume = {81},
  issue = {5},
  pages = {1110--1113},
  numpages = {0},
  year = {1998},
  month = {Aug},
  publisher = {American Physical Society},
  doi = {10.1103/PhysRevLett.81.1110},
  url = {https://link.aps.org/doi/10.1103/PhysRevLett.81.1110}
}

@article{Bulgarini2014,
author = {Bulgarini, Gabriele and Reimer, Michael E. and Bouwes Bavinck, Maaike and J{\"o}ns, Klaus D. and Dalacu, Dan and Poole, Philip J. and Bakkers, Erik P. A. M. and Zwiller, Val},
title = {Nanowire Waveguides Launching Single Photons in a Gaussian Mode for Ideal Fiber Coupling},
journal = {Nano Letters},
volume = {14},
number = {7},
pages = {4102-4106},
year = {2014},
doi = {10.1021/nl501648f},
URL = { 
    
        https://doi.org/10.1021/nl501648f
},

}

@Article{Nikolaev1999,
author={Nikolaev, V. V.
and Sokolovskii, G. S.
and Kaliteevskii, M. A.},
title={Bragg reflectors for cylindrical waves},
journal={Semiconductors},
year={1999},
month={Feb},
day={01},
volume={33},
number={2},
pages={147-152},
issn={1090-6479},
doi={10.1134/1.1187661},
url={https://doi.org/10.1134/1.1187661}
}

@article{Nikolaev1999v2,
author = { M. A.   Kaliteevski  and  R. A.   Abram  and  V. V.   Nikolaev  and  G. S.   Sokolovski },
title = {Bragg reflectors for cylindrical waves},
journal = {Journal of Modern Optics},
volume = {46},
number = {5},
pages = {875-890},
year  = {1999},
publisher = {Taylor & Francis},
doi = {10.1080/09500349908231310},

URL = { 
    
    
        https://www.tandfonline.com/doi/abs/10.1080/09500349908231310
    

},

}

@article{Kaliteevski00,
author = { M. A.   Kaliteevski  and  R. A.   Abram  and  V. V.   Nikolaev },
title = {Optical eigenmodes of a cylindrical microcavity},
journal = {Journal of Modern Optics},
volume = {47},
number = {4},
pages = {677-684},
year  = {2000},
publisher = {Taylor & Francis},
doi = {10.1080/09500340008233388},

URL = { 
    
    
        https://www.tandfonline.com/doi/abs/10.1080/09500340008233388
    

},


}

@article{Kaliteevskii2000,
author={Kaliteevskii, M. A.
and Nikolaev, V. V.},
title={Analogs of the brewster effect and total internal reflection for cylindrical waves},
journal={Technical Physics},
year={2000},
month={Jul},
day={01},
volume={45},
number={7},
pages={865-869},
issn={1090-6525},
doi={10.1134/1.1259740},
url={https://doi.org/10.1134/1.1259740}
}

@article{Jiang94,
author = {Yuan Jiang and Jill Hacker},
journal = {Appl. Opt.},
number = {31},
pages = {7431--7434},
publisher = {Optica Publishing Group},
title = {Cylindrical-wave reflection and antireflection at media interfaces},
volume = {33},
month = {Nov},
year = {1994},
url = {https://opg.optica.org/ao/abstract.cfm?URI=ao-33-31-7431},
doi = {10.1364/AO.33.007431},
}

@article{Jiang93,
author = {Jiang,Yuan  and Hacker,Jill },
title = {Distributed‐{B}ragg reflectors and 90° couplers for cylindrical wave devices},
journal = {Applied Physics Letters},
volume = {63},
number = {11},
pages = {1453-1455},
year = {1993},
doi = {10.1063/1.109653},

URL = {https://doi.org/10.1063/1.109653},
}

@article{Scheuer03v2,
author = {Jacob Scheuer and Amnon Yariv},
journal = {J. Opt. Soc. Am. B},
number = {11},
pages = {2285--2291},
publisher = {Optica Publishing Group},
title = {Annular {B}ragg defect mode resonators},
volume = {20},
month = {Nov},
year = {2003},
url = {https://opg.optica.org/josab/abstract.cfm?URI=josab-20-11-2285},
doi = {10.1364/JOSAB.20.002285},
}

@article{Scheuer03v1,
author = {Jacob Scheuer and Amnon Yariv},
journal = {Opt. Express},
number = {21},
pages = {2736--2746},
publisher = {Optica Publishing Group},
title = {Optical annular resonators based on radial {B}ragg and photonic crystal reflectors},
volume = {11},
month = {Oct},
year = {2003},
url = {https://opg.optica.org/oe/abstract.cfm?URI=oe-11-21-2736},
doi = {10.1364/OE.11.002736},
}

@misc{jiang1994cylindrical,
  title={Cylindrical-wave controlling, generating and guiding devices},
  author={Jiang, Yuan and Hacker, Jill},
  year={1994},
  month=oct # "~18",
  publisher={Google Patents},
  note={US Patent 5,357,591}
}

@article{Ochoa00,
  title = {Diffraction of cylindrical {B}ragg reflectors surrounding an in-plane semiconductor microcavity},
  author = {Ochoa, D. and Houdr\'e, R. and Ilegems, M. and Benisty, H. and Krauss, T. F. and Smith, C. J. M.},
  journal = {Phys. Rev. B},
  volume = {61},
  issue = {7},
  pages = {4806--4812},
  numpages = {0},
  year = {2000},
  month = {Feb},
  publisher = {American Physical Society},
  doi = {10.1103/PhysRevB.61.4806},
  url = {https://link.aps.org/doi/10.1103/PhysRevB.61.4806}
}

@inproceedings{Scheuersensing,
author = {Jacob Scheuer and William M. J. Green and Guy DeRose and Amnon Yariv},
title = {{Annular Bragg defect mode resonators}},
volume = {5333},
booktitle = {Laser Resonators and Beam Control VII},
editor = {Alexis V. Kudryashov},
organization = {International Society for Optics and Photonics},
publisher = {SPIE},
pages = {183 -- 194},
year = {2004},
doi = {10.1117/12.544590},
URL = {https://doi.org/10.1117/12.544590}
}

@article{Scheuerlasing,
author = {Jacob Scheuer and William M. J. Green and Guy DeRose and Amnon Yariv},
journal = {Opt. Lett.},
number = {22},
pages = {2641--2643},
publisher = {Optica Publishing Group},
title = {Low-threshold two-dimensional annular {B}ragg lasers},
volume = {29},
month = {Nov},
year = {2004},
url = {https://opg.optica.org/ol/abstract.cfm?URI=ol-29-22-2641},
doi = {10.1364/OL.29.002641},
}

@ARTICLE{Scheuerlasing3,
  author={Scheuer, J. and Green, W.M.J. and DeRose, G.A. and Yariv, A.},
  journal={IEEE Journal of Selected Topics in Quantum Electronics}, 
  title={In{G}a{A}s{P} annular {B}ragg lasers: theory, applications, and modal properties}, 
  year={2005},
  volume={11},
  number={2},
  pages={476-484},
  doi={10.1109/JSTQE.2005.845614}}

@article{Scheuer:07,
author = {Jacob Scheuer},
journal = {J. Opt. Soc. Am. B},
number = {9},
pages = {2178--2184},
publisher = {Optica Publishing Group},
title = {Radial Bragg lasers: optimal design for minimal threshold levels and enhanced mode discrimination},
volume = {24},
month = {Sep},
year = {2007},
url = {https://opg.optica.org/josab/abstract.cfm?URI=josab-24-9-2178},
doi = {10.1364/JOSAB.24.002178},
}

@article{Scheuer10,
author = {Weiss,Ori  and Scheuer,Jacob },
title = {Emission of cylindrical and elliptical vector beams from radial {B}ragg Lasers},
journal = {Applied Physics Letters},
volume = {97},
number = {25},
pages = {251108},
year = {2010},
doi = {10.1063/1.3529462},

URL = { 
        https://doi.org/10.1063/1.3529462
    
},
}

@INPROCEEDINGS{braggfiber2,
  author={Vienne, G. and Xu, Y. and Jakobsen, C. and Deyerl, H.J. and Hansen, T.P. and Larsen, B.H. and Jensen, J.B. and Sorensen, T. and Terrel, M. and Huang, Y. and Lee, R. and Mortensen, N.A. and Broeng, J. and Simonsen, H. and Bjarklev, A. and Yariv, A.},
  booktitle={Optical Fiber Communication Conference, 2004. OFC 2004}, 
  title={First demonstration of air-silica {B}ragg fiber}, 
  year={2004},
  volume={2},
  number={},
  pages={3 pp. vol.2-},
  doi={}}

@article{Scheuerlaser2,
author = {Scheuer,Jacob  and Green,William M. J.  and DeRose,Guy A.  and Yariv,Amnon },
title = {Lasing from a circular {B}ragg nanocavity with an ultrasmall modal volume},
journal = {Applied Physics Letters},
volume = {86},
number = {25},
pages = {251101},
year = {2005},
doi = {10.1063/1.1947375},

URL = { 
        https://doi.org/10.1063/1.1947375
    
},

}

@article{Johnson:01,
author = {Steven G. Johnson and Mihai Ibanescu and M. Skorobogatiy and Ori Weisberg and Torkel D. Engeness and Marin Solja\v{c}i\'{c} and Steven A. Jacobs and J. D. Joannopoulos and Yoel Fink},
journal = {Opt. Express},
number = {13},
pages = {748--779},
publisher = {Optica Publishing Group},
title = {Low-loss asymptotically single-mode propagation in large-core {O}mni{G}uide fibers},
volume = {9},
month = {Dec},
year = {2001},
url = {https://opg.optica.org/oe/abstract.cfm?URI=oe-9-13-748},
doi = {10.1364/OE.9.000748},
}

@ARTICLE{Micro_ring_FDTD,
  author={Ho, Ying-Lung Daniel and Cao, Tun and Ivanov, Pavel S. and Cryan, Martin J. and Craddock, Ian J. and Railton, Chris J. and Rarity, John G.},
  journal={IEEE Journal of Quantum Electronics}, 
  title={Three-{D}imensional {FDTD} {S}imulation of {M}icro-{P}illar {M}icrocavity {G}eometries {S}uitable for {E}fficient {S}ingle-{P}hoton {S}ources}, 
  year={2007},
  volume={43},
  number={6},
  pages={462-472},
  doi={10.1109/JQE.2007.897905}}

@article{Blokhin:21,
author = {S. A. Blokhin and M. A. Bobrov and N. A. Maleev and J. N. Donges and L. Bremer and A. A. Blokhin and A. P. Vasil'ev and A. G. Kuzmenkov and E. S. Kolodeznyi and V. A. Shchukin and N. N. Ledentsov and S. Reitzenstein and V. M. Ustinov},
journal = {Opt. Express},
number = {5},
pages = {6582--6598},
publisher = {Optica Publishing Group},
title = {Design optimization for bright electrically-driven quantum dot single-photon sources emitting in telecom {O}-band},
volume = {29},
month = {Mar},
year = {2021},
url = {https://opg.optica.org/oe/abstract.cfm?URI=oe-29-5-6582},
doi = {10.1364/OE.415979},
}

@article{Jakubczyk2014,
author = {Jakubczyk, Tomasz and Franke, Helena and Smoleński, Tomasz and Ściesiek, Maciej and Pacuski, Wojciech and Golnik, Andrzej and Schmidt-Grund, Rüdiger and Grundmann, Marius and Kruse, Carsten and Hommel, Detlef and Kossacki, Piotr},
title = {Inhibition and {E}nhancement of the {S}pontaneous {E}mission of {Q}uantum {D}ots in {M}icropillar {C}avities with {R}adial-{D}istributed {B}ragg {R}eflectors},
journal = {ACS Nano},
volume = {8},
number = {10},
pages = {9970-9978},
year = {2014},
doi = {10.1021/nn5017555},
    note ={PMID: 25181393},

URL = { 
        https://doi.org/10.1021/nn5017555
    
},

}

@Article{Manenkov1970,
author={Manenkov, A. B.},
title={The excitation of open homogeneous waveguides},
journal={Radiophysics and Quantum Electronics},
year={1970},
month={May},
day={01},
volume={13},
number={5},
pages={578-586},
issn={1573-9120},
doi={10.1007/BF01030694},
url={https://doi.org/10.1007/BF01030694}
}

@ARTICLE{Snyder1971,
author = {Snyder, Allan W.},
doi = {10.1109/TMTT.1971.1127613},
issn = {15579670},
journal = {IEEE Transactions on Microwave Theory and Techniques},
number = {8},
pages = {720--727},
title = {{Continuous Mode Spectrum of a Circular Dielectric Rod}},
volume = {19},
year = {1971}
}

@phdthesis{SammutPHD,
    title    = {The theory of unbound modes on circular dielectric waveguides},
    school   = {The Australian National University},
    author   = {Sammut, Rowland Alexander},
    year     = {1975},
}

@ARTICLE{Nyquist1981,
author = {Nyquist, Dennis P. and Johnson, Dean R. and Hsu, S. Victor},
doi = {10.1364/JOSA.71.000049},
issn = {00303941},
journal = {J. Opt. Soc. Am.},
number = {1},
pages = {49--54},
title = {{Orthogonality and Amplitude Spectrum of Radiation Modes Along Open-Boundary Waveguides.}},
volume = {71},
year = {1981}
}

@ARTICLE{Vassallo81,
author = {Charles Vassallo},
journal = {J. Opt. Soc. Am.},
number = {10},
pages = {1282--1282},
publisher = {Optica Publishing Group},
title = {Orthogonality and amplitude spectrum of radiation modes along open-boundary waveguides: comment},
volume = {71},
month = {Oct},
year = {1981},
url = {http://opg.optica.org/abstract.cfm?URI=josa-71-10-1282},
doi = {10.1364/JOSA.71.001282},
}

@article{Shevchenko82,
author = {Shevchenko, V. V.},
title = {On the completeness of spectral expansion of the electromagnetic field in the set of dielectric circular rod waveguide eigen waves},
journal = {Radio Science},
volume = {17},
number = {1},
pages = {229-231},
doi = {10.1029/RS017i001p00229},
url = {https://agupubs.onlinelibrary.wiley.com/doi/abs/10.1029/RS017i001p00229},
year = {1982}
}

@ARTICLE{Sammut1982,
author = {Rowland A. Sammut},
journal = {J. Opt. Soc. Am.},
number = {10},
pages = {1335--1337},
publisher = {Optica Publishing Group},
title = {Orthogonality and normalization of radiation modes in dielectric waveguides},
volume = {72},
month = {Oct},
year = {1982},
url = {http://opg.optica.org/abstract.cfm?URI=josa-72-10-1335},
doi = {10.1364/JOSA.72.001335},
}

@ARTICLE{Vassallo83,
author = {C. Vassallo},
journal = {J. Opt. Soc. Am.},
number = {5},
pages = {680--683},
publisher = {Optica Publishing Group},
title = {Orthogonality and normalization of radiation modes in dielectric waveguides: an alternative derivation},
volume = {73},
month = {May},
year = {1983},
url = {http://opg.optica.org/abstract.cfm?URI=josa-73-5-680},
doi = {10.1364/JOSA.73.000680},
}

@book{Snyder1983,
author = {Snyder, Allan W. and Love, John D.},
publisher = {Springer New York, NY},
title = {{Optical waveguide theory}},
year = {1983}
}

@ARTICLE{Tigelis1987,
author = {Tigelis, I. and Capsalis, C. N. and Uzunoglu, N. K.},
doi = {10.1007/BF01010811},
issn = {01959271},
journal = {International Journal of Infrared and Millimeter Waves},
number = {9},
pages = {1053--1068},
title = {{Computation of the dielectric rod waveguide radiation modes}},
volume = {8},
year = {1987}
}

@ARTICLE{Morita1988,
author = {Morita, N.},
doi = {10.1163/156939388X00080},
issn = {15693937},
journal = {Journal of Electromagnetic Waves and Applications},
number = {5-6},
pages = {445--457},
title = {{Radiation Modes of Circular Dielectric Waveguides}},
volume = {2},
year = {1988}
}

@ARTICLE{Alvarez2005,
author = {Alvarez, L. and Xiao, M.},
doi = {10.1163/156939305775468688},
issn = {09205071},
journal = {Journal of Electromagnetic Waves and Applications},
number = {7},
pages = {933--951},
title = {{Complete normalized orthogonal set including evanescent modes for circular dielectric waveguides}},
volume = {19},
year = {2005}
}

@article{Yeh:78,
author = {Pochi Yeh and Amnon Yariv and Emanuel Marom},
journal = {J. Opt. Soc. Am.},
number = {9},
pages = {1196--1201},
publisher = {Optica Publishing Group},
title = {Theory of Bragg fiber$\ast$},
volume = {68},
month = {Sep},
year = {1978},
url = {https://opg.optica.org/abstract.cfm?URI=josa-68-9-1196},
doi = {10.1364/JOSA.68.001196},
}

@article{Jebali:07,
author = {Asma Jebali and Daniel Erni and Stephan Gulde and Rainer F. Mahrt and Werner B\"{a}chtold},
journal = {J. Opt. Soc. Am. B},
number = {4},
pages = {906--915},
publisher = {Optica Publishing Group},
title = {Analytical calculation of the {Q} factor for circular-grating microcavities},
volume = {24},
month = {Apr},
year = {2007},
url = {https://opg.optica.org/josab/abstract.cfm?URI=josab-24-4-906},
doi = {10.1364/JOSAB.24.000906},
}

@article{Claudon13,
author = {Claudon, Julien and Gregersen, Niels and Lalanne, Philippe and Gérard, Jean-Michel},
title = {Harnessing Light with Photonic Nanowires: Fundamentals and Applications to Quantum Optics},
journal = {ChemPhysChem},
volume = {14},
number = {11},
pages = {2393-2402},
doi = {10.1002/cphc.201300033},
url = {https://chemistry-europe.onlinelibrary.wiley.com/doi/abs/10.1002/cphc.201300033},
year = {2013}
}

@article{wang2021,
author = {Wang,Bi-Ying  and Häyrynen,Teppo  and Vannucci,Luca  and Jacobsen,Martin Arentoft  and Lu,Chao-Yang  and Gregersen,Niels },
title = {Suppression of background emission for efficient single-photon generation in micropillar cavities},
journal = {Applied Physics Letters},
volume = {118},
number = {11},
pages = {114003},
year = {2021},
doi = {10.1063/5.0044018},

URL = { 
        https://doi.org/10.1063/5.0044018
    
},

}

@article{Thomas2021,
  title = {Bright Polarized Single-Photon Source Based on a Linear Dipole},
  author = {Thomas, S. E. and Billard, M. and Coste, N. and Wein, S. C. and Priya and Ollivier, H. and Krebs, O. and Taza\"{\i}rt, L. and Harouri, A. and Lemaitre, A. and Sagnes, I. and Anton, C. and Lanco, L. and Somaschi, N. and Loredo, J. C. and Senellart, P.},
  journal = {Phys. Rev. Lett.},
  volume = {126},
  issue = {23},
  pages = {233601},
  numpages = {6},
  year = {2021},
  month = {Jun},
  publisher = {American Physical Society},
  doi = {10.1103/PhysRevLett.126.233601},
  url = {https://link.aps.org/doi/10.1103/PhysRevLett.126.233601}
}

@article{Fons2018,
author = {Fons, Romain and Osterkryger, Andreas D. and Stepanov, Petr and Gautier, Eric and Bleuse, Joël and G{\'e}rard, Jean-Michel and Gregersen, Niels and Claudon, Julien},
title = {{A}ll-{O}ptical {M}apping of the {P}osition of {Q}uantum {D}ots {E}mbedded in a {N}anowire {A}ntenna},
journal = {Nano Letters},
volume = {18},
number = {10},
pages = {6434-6440},
year = {2018},
doi = {10.1021/acs.nanolett.8b02826},

URL = { 
    
        https://doi.org/10.1021/acs.nanolett.8b02826
    
    

},


}

@article{Choi2021,
    author = {Choi, Minho and Kim, Sejeong and Choi, Sunghan and Cho, Yong-Hoon},
    title = {Photonic rocket structure grown by site-selective and bottom-up approach: A directional and {G}aussian-like quantum emitter platform},
    journal = {Applied Physics Letters},
    volume = {119},
    number = {3},
    pages = {034001},
    year = {2021},
    month = {07},
    issn = {0003-6951},
    doi = {10.1063/5.0046084},
    url = {https://doi.org/10.1063/5.0046084},
}

@article{Leandro2018,
author = {Leandro, Lorenzo and Gunnarsson, Christine P. and Reznik, Rodion and J{\"o}ns, Klaus D. and Shtrom, Igor and Khrebtov, Artem and Kasama, Takeshi and Zwiller, Valery and Cirlin, George and Akopian, Nika},
title = {Nanowire Quantum Dots Tuned to Atomic Resonances},
journal = {Nano Letters},
volume = {18},
number = {11},
pages = {7217-7221},
year = {2018},
doi = {10.1021/acs.nanolett.8b03363},

URL = { 
    
        https://doi.org/10.1021/acs.nanolett.8b03363
    
    

},


}

@article{Cadeddu16,
author = {Cadeddu,Davide  and Teissier,Jean  and Braakman,Floris R.  and Gregersen,Niels  and Stepanov,Petr  and Gérard,Jean-Michel  and Claudon,Julien  and Warburton,Richard J.  and Poggio,Martino  and Munsch,Mathieu },
title = {A fiber-coupled quantum-dot on a photonic tip},
journal = {Applied Physics Letters},
volume = {108},
number = {1},
pages = {011112},
year = {2016},
doi = {10.1063/1.4939264},

URL = { 
        https://doi.org/10.1063/1.4939264
    
},

}

@article{Stepanov15,
author = {Stepanov,Petr  and Delga,Adrien  and Gregersen,Niels  and Peinke,Emanuel  and Munsch,Mathieu  and Teissier,Jean  and Mørk,Jesper  and Richard,Maxime  and Bleuse,Joël  and Gérard,Jean-Michel  and Claudon,Julien },
title = {Highly directive and {G}aussian far-field emission from “giant” photonic trumpets},
journal = {Applied Physics Letters},
volume = {107},
number = {14},
pages = {141106},
year = {2015},
doi = {10.1063/1.4932574},

URL = { 
        https://doi.org/10.1063/1.4932574
    
},


}

@article{Gregersen:10,
author = {Niels Gregersen and Torben Roland Nielsen and Jesper M{\o}rk and Julien Claudon and Jean-Michel G\'{e}rard},
journal = {Opt. Express},
number = {20},
pages = {21204--21218},
publisher = {Optica Publishing Group},
title = {Designs for high-efficiency electrically pumped photonic nanowire single-photon sources},
volume = {18},
month = {Sep},
year = {2010},
url = {https://opg.optica.org/oe/abstract.cfm?URI=oe-18-20-21204},
doi = {10.1364/OE.18.021204},
}

@article{Gregersen:08,
author = {Niels Gregersen and Torben R. Nielsen and Julien Claudon and Jean-Michel G\'{e}rard and Jesper M{\o}rk},
journal = {Opt. Lett.},
number = {15},
pages = {1693--1695},
publisher = {Optica Publishing Group},
title = {Controlling the emission profile of a nanowire with a conical taper},
volume = {33},
month = {Aug},
year = {2008},
url = {https://opg.optica.org/ol/abstract.cfm?URI=ol-33-15-1693},
doi = {10.1364/OL.33.001693},
}

@article{ostfeldt22,
  title = {On-Demand Source of Dual-Rail Photon Pairs Based on Chiral Interaction in a Nanophotonic Waveguide},
  author = {\O{}stfeldt, Freja T. and Gonz\'alez-Ruiz, Eva M. and Hauff, Nils and Wang, Ying and Wieck, Andreas D. and Ludwig, Arne and Schott, R\"udiger and Midolo, Leonardo and S\o{}rensen, Anders S. and Uppu, Ravitej and Lodahl, Peter},
  journal = {PRX Quantum},
  volume = {3},
  issue = {2},
  pages = {020363},
  numpages = {10},
  year = {2022},
  month = {Jun},
  publisher = {American Physical Society},
  doi = {10.1103/PRXQuantum.3.020363},
  url = {https://link.aps.org/doi/10.1103/PRXQuantum.3.020363}
}

@Article{Smith2017,
author={Iles-Smith, Jake
and McCutcheon, Dara P. S.
and Nazir, Ahsan
and M{\o}rk, Jesper},
title={Phonon scattering inhibits simultaneous near-unity efficiency and indistinguishability in semiconductor single-photon sources},
journal={Nature Photonics},
year={2017},
month={Aug},
day={01},
volume={11},
number={8},
pages={521-526},
issn={1749-4893},
doi={10.1038/nphoton.2017.101},
url={https://doi.org/10.1038/nphoton.2017.101}
}

@ARTICLE{Doran1983,
  author={Doran, N. and Blow, K.},
  journal={Journal of Lightwave Technology}, 
  title={Cylindrical Bragg fibers: A design and feasibility study for optical communications}, 
  year={1983},
  volume={1},
  number={4},
  pages={588-590},
  doi={10.1109/JLT.1983.1072171}}

@article{Moreau2001,
    author = {Moreau, E. and Robert, I. and Gérard, J. M. and Abram, I. and Manin, L. and Thierry-Mieg, V.},
    title = "{Single-mode solid-state single photon source based on isolated quantum dots in pillar microcavities}",
    journal = {Applied Physics Letters},
    volume = {79},
    number = {18},
    pages = {2865-2867},
    year = {2001},
    month = {10},
    issn = {0003-6951},
    doi = {10.1063/1.1415346},
    url = {https://doi.org/10.1063/1.1415346},
}

@Article{Barnes2002,
author={Barnes, W. L.
and Bj{\"o}rk, G.
and G{\'e}rard, J. M.
and Jonsson, P.
and Wasey, J. A. E.
and Worthing, P. T.
and Zwiller, V.},
title={Solid-state single photon sources: light collection strategies},
journal={The European Physical Journal D - Atomic, Molecular, Optical and Plasma Physics},
year={2002},
month={Feb},
day={01},
volume={18},
number={2},
pages={197-210},
issn={1434-6079},
doi={10.1140/epjd/e20020024},
url={https://doi.org/10.1140/epjd/e20020024}
}

@ARTICLE{Gerard1999,
  author={G\'erard, J.-M. and Gayral, B.},
  journal={Journal of Lightwave Technology}, 
  title={Strong $\mathrm{P}$urcell effect for {I}n{A}s quantum boxes in three-dimensional solid-state microcavities}, 
  year={1999},
  volume={17},
  number={11},
  pages={2089-2095},
  doi={10.1109/50.802999}}

@article{Michler2000,
author = {P. Michler  and A. Kiraz  and C. Becher  and W. V. Schoenfeld  and P. M. Petroff  and Lidong Zhang  and E. Hu  and A. Imamoglu },
title = {A {Q}uantum {D}ot {S}ingle-{P}hoton {T}urnstile {D}evice},
journal = {Science},
volume = {290},
number = {5500},
pages = {2282-2285},
year = {2000},
doi = {10.1126/science.290.5500.2282}}

@Article{Santori2002,
author={Santori, Charles
and Fattal, David
and Vu{\v{c}}kovi{\'{c}}, Jelena
and Solomon, Glenn S.
and Yamamoto, Yoshihisa},
title={Indistinguishable photons from a single-photon device},
journal={Nature},
year={2002},
month={Oct},
day={01},
volume={419},
number={6907},
pages={594-597},
issn={1476-4687},
doi={10.1038/nature01086},
url={https://doi.org/10.1038/nature01086}
}

@article{Madigawa2024,
author = {Madigawa, Abdulmalik A. and Donges, Jan N. and Gaál, Benedek and Li, Shulun and Jacobsen, Martin Arentoft and Liu, Hanqing and Dai, Deyan and Su, Xiangbin and Shang, Xiangjun and Ni, Haiqiao and Schall, Johannes and Rodt, Sven and Niu, Zhichuan and Gregersen, Niels and Reitzenstein, Stephan and Munkhbat, Battulga},
title = {Assessing the Alignment Accuracy of State-of-the-Art Deterministic Fabrication Methods for Single Quantum Dot Devices},
journal = {ACS Photonics},
volume = {11},
number = {3},
pages = {1012-1023},
year = {2024},
doi = {10.1021/acsphotonics.3c01368},

URL = { 
    
        https://doi.org/10.1021/acsphotonics.3c01368
    
    

},


}

@article{Heindel2023,
author = {Heindel, Tobias and Kim, Je-Hyung and Gregersen, Niels and Rastelli, Armando and Reitzenstein, Stephan},
doi = {10.1364/AOP.490091},
issn = {1943-8206},
journal = {Adv. Opt. Photonics},
month = {sep},
number = {3},
pages = {613--738},
publisher = {Optica Publishing Group},
title = {{Quantum dots for photonic quantum information technology}},
url = {https://opg.optica.org/viewmedia.cfm?uri=aop-15-3-613{\&}seq=0{\&}html=true https://opg.optica.org/abstract.cfm?uri=aop-15-3-613 https://opg.optica.org/aop/abstract.cfm?uri=aop-15-3-613},
volume = {15},
year = {2023}
}

@article{Schneider:18,
author = {Philipp-Immanuel Schneider and Nicole Srocka and Sven Rodt and Lin Zschiedrich and Stephan Reitzenstein and Sven Burger},
journal = {Opt. Express},
number = {7},
pages = {8479--8492},
publisher = {Optica Publishing Group},
title = {Numerical optimization of the extraction efficiency of a quantum-dot based single-photon emitter into a single-mode fiber},
volume = {26},
month = {Apr},
year = {2018},
url = {https://opg.optica.org/oe/abstract.cfm?URI=oe-26-7-8479},
doi = {10.1364/OE.26.008479},
}

@article{Versteegh2014,
author={Versteegh, Marijn A. M.
and Reimer, Michael E.
and J{\"o}ns, Klaus D.
and Dalacu, Dan
and Poole, Philip J.
and Gulinatti, Angelo
and Giudice, Andrea
and Zwiller, Val},
title={Observation of strongly entangled photon pairs from a nanowire quantum dot},
journal={Nature Communications},
year={2014},
month={Oct},
day={31},
volume={5},
number={1},
pages={5298},
issn={2041-1723},
doi={10.1038/ncomms6298},
url={https://doi.org/10.1038/ncomms6298}
}

@article{gines2022,
  title = {High Extraction Efficiency Source of Photon Pairs Based on a Quantum Dot Embedded in a Broadband Micropillar Cavity},
  author = {Gin\'es, Laia and Mocza\l{}a-Dusanowska, Magdalena and Dlaka, David and Ho\ifmmode \check{s}\else \v{s}\fi{}\'ak, Radim and Gonzales-Ureta, Junior R. and Lee, Jaewon and Je\ifmmode \check{z}\else \v{z}\fi{}ek, Miroslav and Harbord, Edmund and Oulton, Ruth and H\"ofling, Sven and Young, Andrew B. and Schneider, Christian and Predojevi\ifmmode \acute{c}\else \'{c}\fi{}, Ana},
  journal = {Phys. Rev. Lett.},
  volume = {129},
  issue = {3},
  pages = {033601},
  numpages = {6},
  year = {2022},
  month = {Jul},
  publisher = {American Physical Society},
  doi = {10.1103/PhysRevLett.129.033601},
  url = {https://link.aps.org/doi/10.1103/PhysRevLett.129.033601}
}

@article{Thyrrestrup2010,
    author = {Thyrrestrup, Henri and Sapienza, Luca and Lodahl, Peter},
    title = {Extraction of the $\beta$-factor for single quantum dots coupled to a photonic crystal waveguide},
    journal = {Applied Physics Letters},
    volume = {96},
    number = {23},
    pages = {231106},
    year = {2010},
    month = {06},
    issn = {0003-6951},
    doi = {10.1063/1.3446873},
    url = {https://doi.org/10.1063/1.3446873},
}

@article{Denning:20,
author = {Emil V. Denning and Jake Iles-Smith and Niels Gregersen and Jesper Mork},
journal = {Opt. Mater. Express},
number = {1},
pages = {222--239},
publisher = {Optica Publishing Group},
title = {Phonon effects in quantum dot single-photon sources},
volume = {10},
month = {Jan},
year = {2020},
url = {https://opg.optica.org/ome/abstract.cfm?URI=ome-10-1-222},
doi = {10.1364/OME.380601},
}

@article{Huber_2015,
doi = {10.1088/1367-2630/17/12/123025},
url = {https://doi.org/10.1088/1367-2630/17/12/123025},
year = {2015},
month = {dec},
publisher = {IOP Publishing},
volume = {17},
number = {12},
pages = {123025},
author = {Huber, Tobias and Predojević, Ana and Föger, Daniel and Solomon, Glenn and Weihs, Gregor},
title = {Optimal excitation conditions for indistinguishable photons from quantum dots},
journal = {New Journal of Physics}
}

@article{Karli2022,
author = {Karli, Yusuf and Kappe, Florian and Remesh, Vikas and Bracht, Thomas K. and M{\"u}nzberg, Julian and Covre da Silva, Saimon and Seidelmann, Tim and Axt, Vollrath Martin and Rastelli, Armando and Reiter, Doris E. and Weihs, Gregor},
title = {SUPER Scheme in Action: Experimental Demonstration of Red-Detuned Excitation of a Quantum Emitter},
journal = {Nano Letters},
volume = {22},
number = {16},
pages = {6567-6572},
year = {2022},
doi = {10.1021/acs.nanolett.2c01783},
URL = { 
    
        https://doi.org/10.1021/acs.nanolett.2c01783
    
    

},

}

@misc{Piccinini2025,
      title={Exciton and biexciton preparation via coherent swing-up excitation in a {G}a{A}s quantum dot embedded in micropillar cavity}, 
      author={Claudia Piccinini and Aleksander Rodek and Abdulmalik A. Madigawa and Ailton Garcia Jr. and Saimon F. Covre da Silva and Martin A. Jacobsen and Luca Vannucci and Gregor Weihs and Armando Rastelli and Vikas Remesh and Niels Gregersen and Battulga Munkhbat},
      year={2025},
      eprint={2510.21428},
      archivePrefix={arXiv},
      primaryClass={physics.optics},
      url={https://arxiv.org/abs/2510.21428}, 
}

@article{Sbresny2022,
  title = {Stimulated Generation of Indistinguishable Single Photons from a Quantum Ladder System},
  author = {Sbresny, Friedrich and Hanschke, Lukas and Sch\"oll, Eva and Rauhaus, William and Scaparra, Bianca and Boos, Katarina and Zubizarreta Casalengua, Eduardo and Riedl, Hubert and del Valle, Elena and Finley, Jonathan J. and J\"ons, Klaus D. and M\"uller, Kai},
  journal = {Phys. Rev. Lett.},
  volume = {128},
  issue = {9},
  pages = {093603},
  numpages = {7},
  year = {2022},
  month = {Mar},
  publisher = {American Physical Society},
  doi = {10.1103/PhysRevLett.128.093603},
  url = {https://link.aps.org/doi/10.1103/PhysRevLett.128.093603}
}

@article{Wei2022,
author={Wei, Yuming
and Liu, Shunfa
and Li, Xueshi
and Yu, Ying
and Su, Xiangbin
and Li, Shulun
and Shang, Xiangjun
and Liu, Hanqing
and Hao, Huiming
and Ni, Haiqiao
and Yu, Siyuan
and Niu, Zhichuan
and Iles-Smith, Jake
and Liu, Jin
and Wang, Xuehua},
title={Tailoring solid-state single-photon sources with stimulated emissions},
journal={Nature Nanotechnology},
year={2022},
month={May},
day={01},
volume={17},
number={5},
pages={470-476},
issn={1748-3395},
doi={10.1038/s41565-022-01092-6},
url={https://doi.org/10.1038/s41565-022-01092-6}
}

@article{Yan2022,
author = {Yan, Junyong and Liu, Shunfa and Lin, Xing and Ye, Yongzheng and Yu, Jiawang and Wang, Lingfang and Yu, Ying and Zhao, Yanhui and Meng, Yun and Hu, Xiaolong and Wang, Da-Wei and Jin, Chaoyuan and Liu, Feng},
title = {Double-Pulse Generation of Indistinguishable Single Photons with Optically Controlled Polarization},
journal = {Nano Letters},
volume = {22},
number = {4},
pages = {1483-1490},
year = {2022},
doi = {10.1021/acs.nanolett.1c03543},
URL = { 
    
        https://doi.org/10.1021/acs.nanolett.1c03543
    
    

},
}

@article{Margaria2025,
author={Margaria, Nico
and Pastier, Florian
and Bennour, Thinhinane
and Billard, Marie
and Ivanov, Edouard
and Hease, William
and Stepanov, Petr
and Adiyatullin, Albert F.
and Singla, Raksha
and Pont, Mathias
and Descampeaux, Maxime
and Bernard, Alice
and Pishchagin, Anton
and Morassi, Martina
and Lema{\^i}tre, Aristide
and Volz, Thomas
and Giesz, Val{\'e}rian
and Somaschi, Niccolo
and Maring, Nicolas
and Boissier, S{\'e}bastien
and Au, Thi Huong
and Senellart, Pascale},
title={Efficient fibre-pigtailed source of indistinguishable single photons},
journal={Nature Communications},
year={2025},
month={Aug},
day={14},
volume={16},
number={1},
pages={7553},
issn={2041-1723},
doi={10.1038/s41467-025-62712-y},
url={https://doi.org/10.1038/s41467-025-62712-y}
}

@article{Rickert2021,
    author = {Rickert, L. and Schröder, F. and Gao, T. and Schneider, C. and Höfling, S. and Heindel, T.},
    title = {Fiber-pigtailing quantum-dot cavity-enhanced light emitting diodes},
    journal = {Applied Physics Letters},
    volume = {119},
    number = {13},
    pages = {131104},
    year = {2021},
    month = {09},
    issn = {0003-6951},
    doi = {10.1063/5.0063697},
    url = {https://doi.org/10.1063/5.0063697},
}

\section*{Acknowledgements}
This work is funded by the European Research Council (ERC-CoG “UNITY,” Grant No. 865230), the French National Research Agency (Grant No. ANR-19-CE47-0009-02), the European
Union’s Horizon 2020 Research and Innovation Programme under the Marie Skłodowska-Curie Grant (Agreement No. 861097), the Independent Research Fund Denmark (Grant No. DFF-9041-00046B), and the European Union’s Horizon Europe research and innovation programme under EPIQUE project (Grant Agreement No. 101135288).

\section*{Author contributions statement}
N.G., J.C., and J.-M.G. formulated the project. M.A.J.\ developed the analytical method and performed the simulations. M.A.J.\ wrote the manuscript with comments and editing from all authors. N.G.\ and L.V.\ supervised the project. 

\section*{Competing interests}
The authors declare no financial or non-financial competing interests.



\label{LastMainPage}
\includepdf[pages=-]{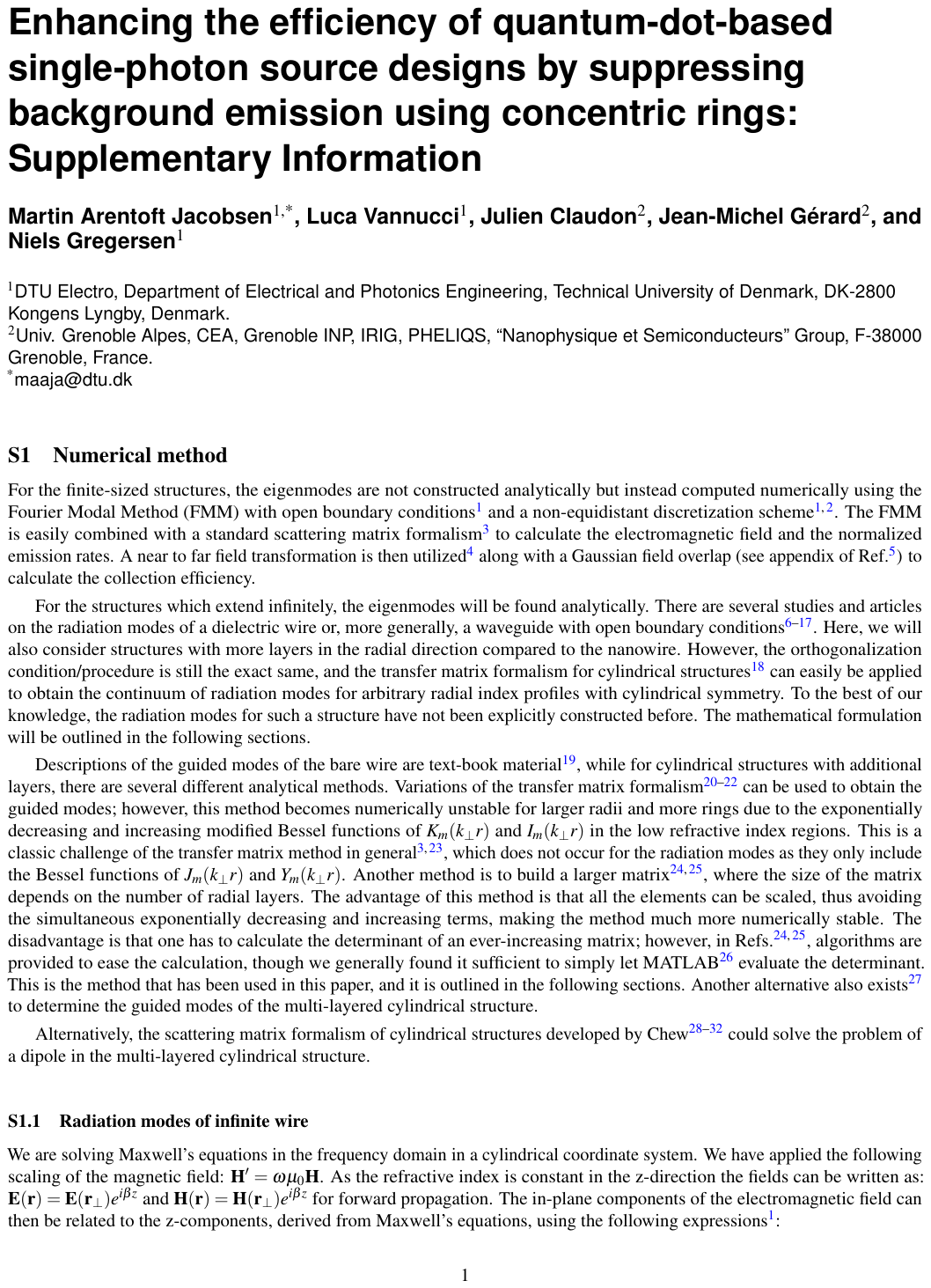}
\end{document}